\documentclass[12pt]{article}

\usepackage{arxiv}
\usepackage{bigfoot}
\usepackage{etoolbox}
\usepackage{layout}
\usepackage{authblk}
\usepackage{hyperref}
\usepackage{amssymb}
\usepackage{amsmath}
\usepackage{amsthm}
\usepackage{lineno}
\usepackage{xcolor}
\usepackage{bbding}
\usepackage{siunitx}
\usepackage{subcaption}
\usepackage{booktabs}
\usepackage{caption}
\usepackage{subfloat}
\usepackage{graphicx}
\usepackage{comment}

\newcommand{\gn}{g_\mathrm{n}}
\newcommand{\gtg}{g_\mathrm{t}}
\newcommand{\pn}{p_\mathrm{n}}
\newcommand{\up}{\partial u(x,t)/ \partial x}
\newcommand{\bb}{b_\mathrm{b}}

\newcommand{\hb}{h_\mathrm{b}}
\newcommand{\en}{\varepsilon_\mathrm{n}}

\newcommand{\Omx}{\Omega(x_1,x_2)}
\newcommand{\Oy}{O(y_1,y_2)}
\newcommand{\yo}{\mathbf{y}_\mathrm{o}(t) = [y_1(t),\,y_2(t)]^\intercal}

\newcommand{\tf}{t_\mathrm{f}}
\newcommand{\yp}{\mathbf{y}_\mathrm{p}=[\mathbf{y}_{\mathrm{p}1},\,\mathbf{y}_{\mathrm{p}2}]^\intercal}
\newcommand{\xp}{\mathbf{x}_\mathrm{p}(t) = \mathbf{y}_\mathrm{o}(t) + \mathbf{y}_\mathrm{p}}

\newcommand{\omu}{\overline{\mu}}
\newcommand{\vmin}{v_\mathrm{min}}
\newcommand{\vmax}{v_\mathrm{max}}

\newcommand{\Pa}{P_\mathrm{a}}
\newcommand{\Pq}{P_\mathrm{q}}
\newcommand{\Pt}{P_\mathrm{t}}
\newcommand{\Ppt}{P_\mathrm{pt}}
\newcommand{\Pvt}{P_\mathrm{vt}}
\newcommand{\Pz}{P_\mathrm{z}}
\newcommand{\Psm}{P_\mathrm{sm}}
\newcommand{\Psmx}{P_\mathrm{smx}}
\newcommand{\Pc}{P_\mathrm{c}}
\newcommand{\Pcx}{P_\mathrm{cx}}

\DeclareNewFootnote{AAffil}[arabic]
\DeclareNewFootnote{ANote}[fnsymbol]

\makeatletter
\patchcmd\maketitle{\def\@makefnmark{\rlap{\@textsuperscript{\normalfont\@thefnmark}}}}{}{}{}
\makeatother

\makeatletter
\def\thanksAAffil#1{
  \footnotemarkAAffil\protected@xdef\@thanks{\@thanks%
        \protect\footnotetextAAffil[\the \c@footnoteAAffil]{#1}}%
}
\def\thanksANote#1{%
  \footnotemarkANote%
  \protected@xdef\@thanks{\@thanks%
        \protect\footnotetextANote[\the \c@footnoteANote]{#1}}%
}
\makeatother

\graphicspath{{figures/}}

\title{A computational framework for evaluating tire-asphalt hysteretic friction including pavement roughness}

\newbox{\orcid}\sbox{\orcid}{\includegraphics[scale=0.06]{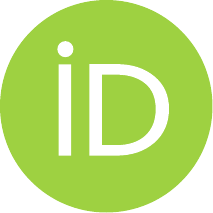}} 
\author[1]{%
	\href{https://orcid.org/0000-0003-0684-3490}{\usebox{\orcid}\hspace{1mm}\textbf{Ivana Ban}\thanks{\texttt{ivana.ban@gradri.uniri.hr}}}%
}
\author[2,3]{%
	\href{https://orcid.org/0000-0001-8435-6466}{\usebox{\orcid}\hspace{1mm}\textbf{Jacopo Bonari}\thanks{\texttt{jacopo.bonari@dlr.de}, corresponding author}}%
}
\author[3]{%
	\href{https://orcid.org/0000-0001-9409-9782}{\usebox{\orcid}\hspace{1mm}\textbf{Marco Paggi}\thanks{\texttt{marco.paggi@imtlucca.it}}}%
}

\affil[1]{Faculty of Civil Engineering, University of Rijeka, Trg Bra\'{c}e Ma\v{z}urani\'{c}a 10, 51000 Rijeka, Croatia}

\affil[2]{Institute for the Protection of Terrestrial Infrastructures, German Aerospace Center (DLR), Rathausallee 12, 53757 Sankt Augustin, Germany}

\affil[3]{IMT School for Advanced Studies Lucca, Piazza San Francesco 19, 56100 Lucca, Italy}

\begin{document}

\maketitle

\begin{abstract}
Pavement surface textures obtained by a photogrammetry-based method for data acquisition and analysis are employed to investigate if related roughness descriptors are comparable to the frictional performance evaluated by finite element analysis. Pavement surface profiles are obtained from 3D digital surface models created with Close-Range Orthogonal Photogrammetry. To characterize the roughness features of analyzed profiles, selected texture parameters were calculated from the profile's geometry.
The parameters' values were compared to the frictional performance obtained by numerical simulations. Contact simulations are performed according to a dedicated finite element scheme where surface roughness is directly embedded into a special class of interface finite elements. Simulations were performed for different case scenarios and the obtained results showed a notable trend between roughness descriptors and friction performance, indicating a promising potential for this numerical method to be consistently employed to predict the frictional properties of actual pavement surface profiles.
\end{abstract}



\section{Introduction}\label{sec:intro}
Pavement surface friction is one of the key functional properties that guarantees safe driving conditions. It affects the contact interface between vehicle tire and asphalt pavement, providing vehicle stability while steering and grip during the braking maneuver~\cite{hall:2009}. It is quantified by the coefficient of friction, a dimensionless number indicating the ratio between the tangential and the normal force. An important feature in friction phenomenon is the limited portion of the interface where the contact actually takes place, a quantity highly affected by the roughness properties of the surface~\cite{bowden:1950}. A close look at most of natural and artificial surfaces reveals a rough topology, that determines a true contact area smaller than it would be experienced for an idealized smooth contact. This feature directly affects the value of the resulting friction coefficient~\cite{Popov:2010}. 

Asphalt pavement surface manifests roughness at many scales of observation, each characterized by several different scales of specific amplitude and wavelength, which in turn are connected to functional pavement properties~\cite{hall:2009}. The macro scale of pavement surface texture, or macro-texture, is usually defined within an amplitude ranging from $0.1\,\si{\milli\meter}$ to $20\,\si{\milli\meter}$ and a wavelength ranging from $0.05\,\si{\milli\meter}$ to $50\,\si{\milli\meter}$. This texture scale is relevant for tire-asphalt interaction and is considered as the governing parameter for functional properties like friction for speeds above $50\,\si{\kilo\meter/\hour}$, aquaplaning, noise, and splash or spray effects. Micro-texture scale is the smallest texture scale for pavements, with defined amplitude limits of $0.001\,\si{\milli\meter}$ to $0.05\,\si{\milli\meter}$ and wavelengths below $0.5\,\si{\milli\meter}$. Micro-texture results from the properties of the aggregate grain material and its morphology is responsible for low speed friction and vehicle tire wear~\cite{hall:2009}. 

An important consideration when dealing with pavement friction is related to the material properties of the bodies in contact. 
For the inelastic materials employed in tires production, rubber friction theory recognizes that the frictional force results from two distinct phenomena, adhesion and hysteresis, that make a different contribution whose combination results in the overall frictional response. Adhesion results from inter-molecular contact between different materials through Van der Waals dipole forces~\cite{yu:2004,sauer:2009}, and is highly dependent on the true contact area size. Hysteresis is a loss of energy that takes place in rubber materials during deformation and recovery. When such a material is stretched and then released, it does not immediately return to its original shape; instead, some of the energy is dissipated as heat due to internal friction within its polymer chains. The energy loss determines a lag between the applied force and the rubber’s response that in turn causes a significant change in the interface contact forces, finally leading to the arousal of tangential forces opposing the sliding motion~\cite{wriggers:2009}. Important contributors to hysteretic friction are sliding speed and rubber temperature~\cite{hall:2009}. 



The non-conforming rough interface between rigid pavement surface and tire rubber makes the true contact area a-priori unknown, and its value varies during the sliding motion, which renders the problem of rubber sliding over a rough profile as highly non-linear~\cite{paggi&hills:2020}. Given the problem complexity, no closed form expressions can be derived to predict the rubber coefficient of friction. To tackle the problem, numerical methods are usually exploited to model rough contact problems including friction. In the research community, two wide spread and well-established methods are usually exploited: the finite element method (FEM)~\cite{zienkiewicz:2002} and the boundary element method (BEM)~\cite{brebbia:2017}, each characterized by their own strengths and weaknesses. FEM requires discretization of the whole domain, including the interior, while BEM, only requires discretization of the boundaries of the domain, thus reducing the dimensionality of the problem. BEM is often more efficient for problems with infinite or semi-infinite domains, since it reduces the number of elements needed. On the other hand, BEM is mostly restricted to simple material laws and geometries, mostly in a small displacement setting. Quite the opposite, FEM is better suited for complex, non-linear, or multi-material problems because it can easily handle a variety of diverse material properties and complex geometries, though more computationally demanding.



Given its characteristics, numerical solutions of pavement-tire interaction problems in the context of friction analysis are most commonly observed within a FEM framework. The physical problem is simulated by modeling the contact between vehicle tires or vehicle tires subsets and the pavement structure~\cite{peng:2019}. Modeling of vehicle tire load requires the discretization of the full tire or of a representative sample, the definition of the specific viscoelastic material model, and often includes anisotropic effects due to the different stiffness of tire constituent materials, i.e., tire rubber and inner embedded steel belt, and specification of loading scenario~\cite{yu:2020}. Pavement structure modeling requires the discretization of both surface and bulk, which is more complex if surface roughness is accounted in its actual form and not simplified. Furthermore, asphalt pavement is characterized as well by a viscoelastic rheology and even though it is mostly considered rigid in comparison to the tire, an adequate material constitutive law also for the pavement can be taken into account in pursuit of a more realistic modeling approach~\cite{yu:2017}.

In the course of the years, FEM analysis of tire-pavement interaction in terms of frictional behavior has been investigated from different research angles. The FEM framework was exploited for the prediction of friction performance to derive the so-called \emph{NUS} skid resistance model~\cite{fwa:2017}. The NUS model is a theoretical model for pavement skid resistance which uses solid mechanics and hydrodynamic theories developed by National University of Singapore (NUS) research group. The model solved the coupled tire-fluid-pavement interaction problem assuming a Coulomb friction law to model the frictional behavior of the system and the Navier-Stokes equations equipped with a standard $k-\varepsilon$ turbulent model to describe the fluid dynamics. However, the work does not consider the influence of pavement surface morphology on the resultant forces, but rather only the influence of the water film thickness and applied sliding speed.
Concurrently,~\cite{srirangam:2017} analyzed the influence of different pavement surface morphological properties on the frictional response of the system. They performed several FEM simulations applying different loading pressures and sliding velocities on three different types of pavements. Excerpts of vehicle tires were simulated as a rectangular viscoelastic rubber block, while the pavement structure was modeled as both rigid or deformable, in either case with a fully discretized texture. Contact was enforced using a surface-to-surface contact algorithm and finite displacements were considered. The sliding response was analyzed through the calculation of a friction coefficient as the ratio between the resultant applied tangential and normal loads.
Research performed by~\cite{yu:2017} focused on the influence of different elastic properties (recoverable resilient deformation) of various pavement structures on the braking performance by setting up a 3D tire-pavement interaction model. They investigated the influence of different system variables such as tread deformation at the contacting interface, actual contact area and the braking force applied within the dynamic friction contact analysis, concluding that pavement elasticity and deformation influence the real contact area. However, they did not study the effect of surface texture on the frictional response of the system. 
Authors~\cite{wagner:2017} performed a multi-scale FEM analysis to calculate low-speed sliding friction on rough rigid pavement surfaces, emphasizing the utmost importance of the two rubber friction components, i.e., adhesion and hysteresis. The simulations were performed in 2D to reduce the computational complexity caused by including in the model real rough surfaces representations and a scale separation hypothesis was applied to account for the effects of micro and macro texture on the frictional response. Hysteretic friction component was accounted by using a viscoelastic material model and a robust surface-to-surface contact algorithm, while the adhesive contribution was added at the macroscopic level only tailored for different surface contact conditions based on a tuning coefficient $\alpha$, whose value ranged from zero to one depending on the surface conditions (zero for surface covered in soap-water mixture and one for dry surface). 
In~\cite{tang:2018}, tire-pavement interaction was investigated within a 3D FEM framework, by emphasizing the effect of high temperatures on the frictional response of the model. Pavement structure was scanned by x-ray tomography and processed to distinguish mixture phases, i.e., aggregates, bitumen, and air voids, and then modeled using micro-structure FEM meshes where aggregates were described as elastic while binder was supposed to be viscoelastic. The authors then modeled the contact behavior using, again, a surface-to-surface contact algorithm and calculated the frictional response from the theory of hysteresis induced energy dissipation. The numerical results were compared to the field measurements of skid resistance showing promising and comparable results.
Research done by~\cite{peng:2019} focused on FEM simulations of frictional response using 3D pavement surface model reconstructed from high resolution pavement texture data in order to explore the influence of texture characteristics on interface friction parameters. They assumed that the pavement was non-deformable, and the rubber characterized as a hyperelastic and viscoelastic block. They assumed an exponential decay friction model proposed by~\cite{oden:1985} taking place at the interface and calculated the friction coefficient from the resultant tangential force over normal applied load. In the study, they explored the water effect on frictional response by lowering the exponential decay friction to mimic the presence of a thin lubricant layer at the interface. The friction model parameters were determined with a binary search back-calculation algorithm, which adjusted the parameters of the FEM simulation such that the simulated and measured skid resistance values resulted comparable.
In~\cite{liu:2019} an integrated tire-vehicle model for the prediction of frictional performance of wet pavements was proposed. They exploited Persson’s friction theory~\cite{persson:1997} for the calculation of the kinetic friction coefficient, based only on the pavement surface power spectra and the rubber rheology specifications. Furthermore, they performed a FEM analysis of the hydroplaning effect to calculate the hydrodynamic forces acting, and determined the effect of presence of water on different tire movements. The characteristics of tire movements were obtained by integrating the calculated friction coefficient into the FEM hydroplaning model and further used for the vehicle dynamics simulation in a dedicated software. This approach considered the influence of pavement texture on the frictional performance, but also linked the frictional response to different vehicle movements while emphasizing the importance of a high friction coefficient during braking or steering maneuvers.

In this research, a FEM approach is emploiyed to simulate the frictional response of tire-pavement interaction considering real pavement texture roughness features in input. To account for the actual geometry of the pavement surface involved, the novel \emph{MPJR} (eMbedded Profile for Joint Roughness) approach is utilized. The approach enables to dramatically simplify the modeling procedure usually required in the definition of a FEM problem involving contacting surface, without introducing geometrical approximations or smoothing. The solution was first introduced in~\cite{paggi:2018,paggi&hills:2020} as an approach to address rough contact problems involving a rigid and a deformable body. The method introduces a novel zero-thickness interface finite element which stores the actual information describing the interface geometry and encodes roughness information directly at each element quadrature nodes, thus allowing to replace the actual complex interface geometry with an equivalent smooth interface, while retaining all the problem's original characteristics.

The method proved to be accurate in the solution of contact problems involving complex interfaces described by analytical expressions and has been further developed to also account for friction in presence of microscopic sliding~\cite{bonari:2021}, finite frictional sliding of harmonic profiles over viscoelastic bodies~\cite{bonari:2020} and adhesive contact problems, both frictionless and with friction~\cite{bonari:2022}. In an ongoing study the method has also been employed to reproduce the results of a contact mechanics challenge~\cite{mueser:2017}, a demanding test case in which researchers in the field of tribology were invited to solve the contact problem between a synthetic large-scale and high-resolution rigid rough surface and an elastic half-space. The MPJR method shows a high degree of accuracy as compared to the benchmark solution provided by the proposers of the challenge.

In the present study, the MPJR framework is further extended to analyze finite sliding scenarios simulating the sliding of viscoelastic bodies over complex rough profiles, whose geometry is directly extracted from 3D pavement surface models obtained with a photogrammetry-based method developed and verified in~\cite{ban:2023}. The object of the study is to test and investigate the possibility of employing the MPJR approach for the evaluation of the hysteretic coefficient of friction, and to draw relationships between the parameter obtained through numerical simulations, describing the frictional performance of the rough profiles, and some selected roughness parameters obtained from the same profiles employed in the numerical simulations. A positive outcome of this preliminary investigation would entitle the proposed method as a valid investigation tool to be consistently employed in this field of research, and would provide motivation for further extended analysis capabilities including, but not limited to, large scale 3D and finite deformation.


\section{Rough profiles description}\label{sec:roughness}

To obtain surface profiles, a novel method for experimental pavement texture data acquisition and analysis is utilized. The method is based on photogrammetry as a technique for object reconstruction from captured digital images~\cite{luhmann:2006}. A special case of photogrammetry employed in this research is close-range photogrammetry, where objects of interest are captured from a close distance to investigate the morphology of the surfaces under examination. To obtain the surface images, the digital camera was mounted $50\,\si{\centi\meter}$ above the pavement surface with camera lens parallel to the surface plane. The method was christened Close-Range Orthogonal Photogrammetry or \emph{CROP} method. The development of CROP method and its verification procedure are thoroughly described in the doctoral thesis~\cite{ban:2023}, while some important highlights are summarized here:
\begin{enumerate}   
    \item Images were captured with a single Nikon D500 DSLR 20.1 Mpix digital camera, with AF Nikkor $50\,\si{\milli\meter}$ f1.8 D single lens. The camera was mounted on a tripod at a fixed height of $50\,\si{\centi\meter}$ and the camera lens positioned parallel to the pavement surface.
    \item The surface of interest was $50\,\si{\centi\meter}\times50\,\si{\centi\meter}$ large, marked by a custom made reference frame for precise digital surface reconstruction and calibration. The frame was created for the method's verification procedure to determine the deviations of dimensions in digital surface model crated by CROP method and true object's dimensions.
    \item Each surface was photographed twenty-five times, with consecutive images overlap equal to $60\%$ side and $80\%$ forward overlap. The camera was moved in parallel rows with respect to the markings on the reference frame, where, in each row, five images were captured.
    \item Images were captured in \texttt{.raw} format to preserve image information in the original form without compression and stored in \texttt{.tiff} format after brightness and contrast optimization, with pixel size $4.45\,\si{\micro\meter}\times4.45\,\si{\micro\meter}$.
    \item Acquired surface images were used for reconstructing a digital surface model in the specialized photogrammetry software Agisoft Metashape v1.5 Pro. Surface reconstruction procedure was done automatically, where captured images and reconstruction parameters were user-defined inputs. The workflow consisted of image alignment based on seek-and-match procedure of common points in captured images, resulting in a sparse point cloud (SPC) entity. The SPC object served as a basis for 3D object reconstruction, subjected to filtering procedure by selected error reduction features: reprojection error, reconstruction uncertainty and projection accuracy. The purpose of error reduction features is to remove the outlier points in captured images based on camera geometry of the images. In this way, the initial SPC entity is filtered and the best-fit tie points are extracted for further reconstruction procedure. The threshold values of error reduction features are specified in~\cite{ban:2023}. The best-fit points filtered from SPC were further used in dense point cloud (DPC) object creation, consisting of a finite number of points with $XYZ$ coordinates, describing the object's geometry and surface morphology. The DPC can be the final object in the reconstruction procedure or it can be used for the reconstruction of 3D mesh object.
\end{enumerate}

\begin{figure}[t]
    \centering
    \subfloat[][Pavement surface data acquisition by CROP method, with orthogonal positioned digital camera and target surface marked with reference frame.\label{fig:crop}]
        {\includegraphics[height=0.27\textwidth]{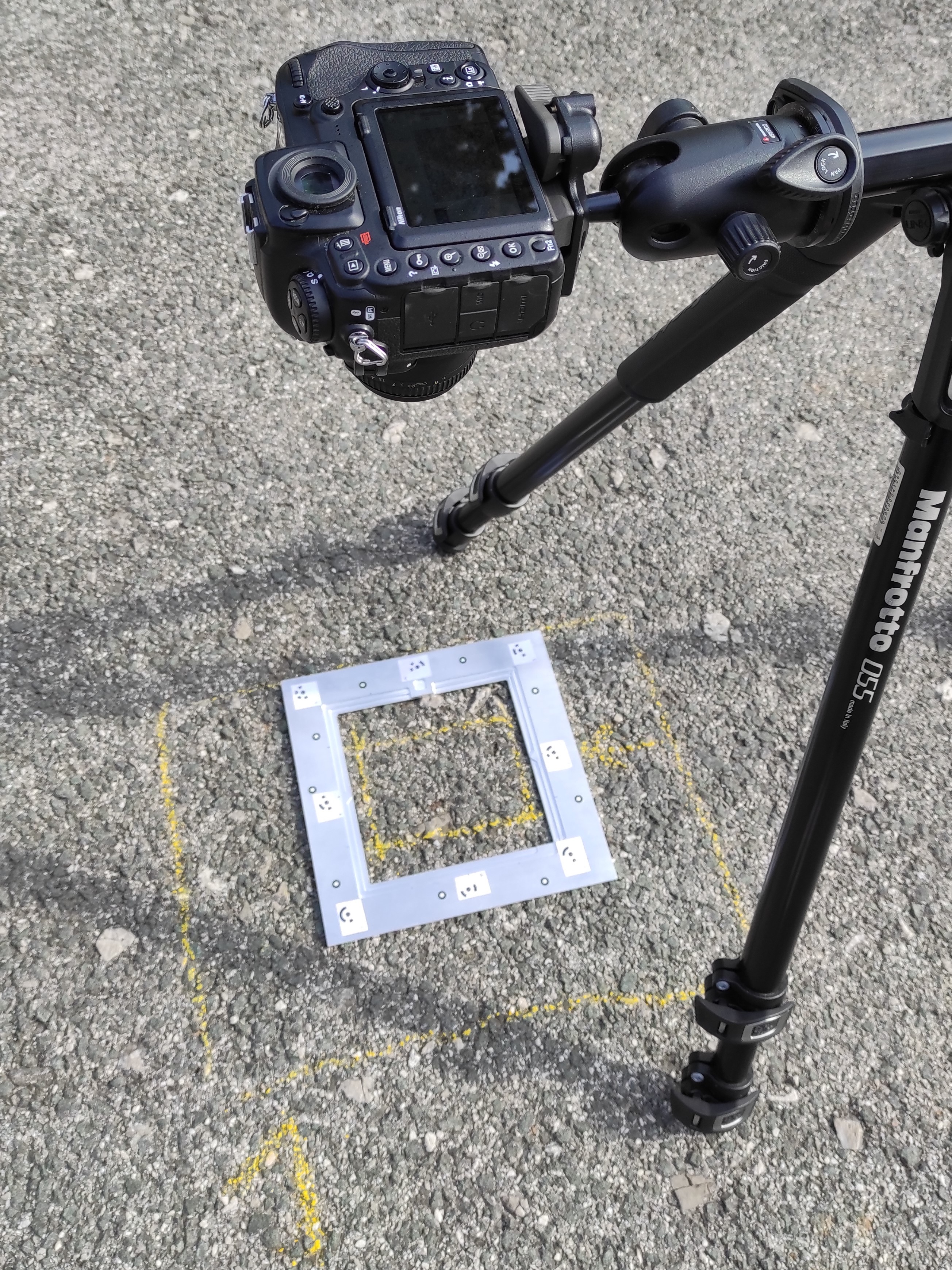}}\hspace{5mm}%
    \subfloat[][Digital surface model of pavement surface obtained by CROP method application.\label{fig:surface_crop}]
        {\includegraphics[height=0.27\textwidth]{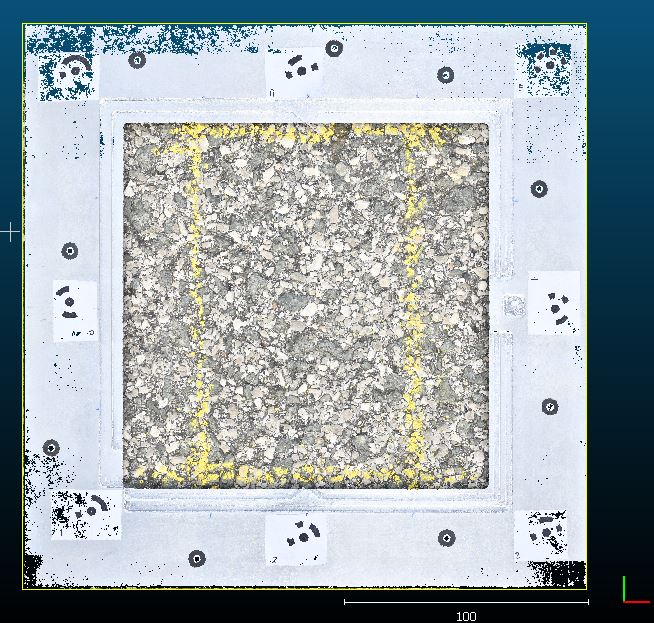}}\hspace{5mm}%
    \subfloat[][An example of digital pavement surface model processed in the Cloud Compare software \label{fig:surface_digital}]
        {\includegraphics[height=0.27\textwidth]{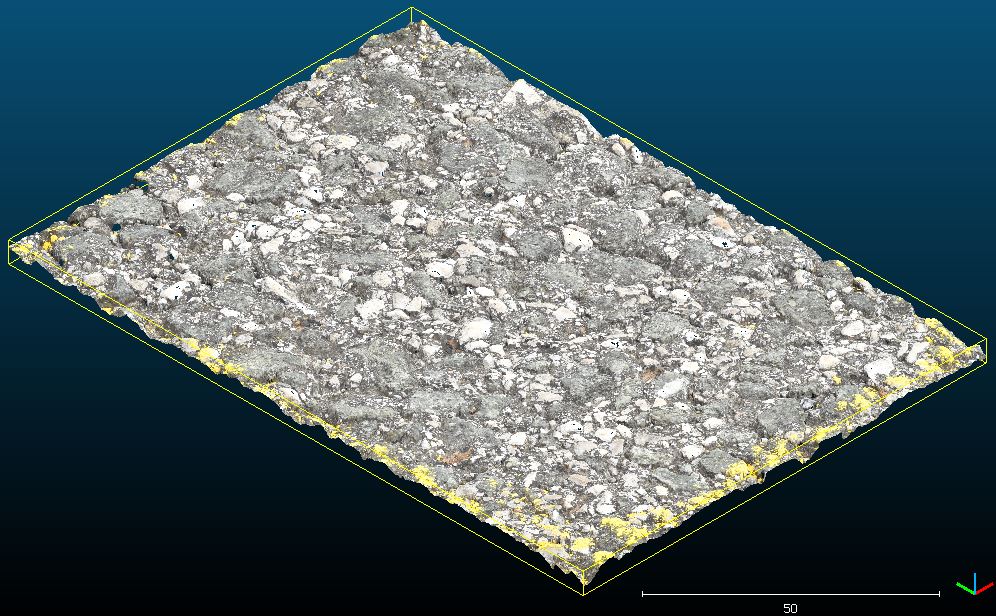}}
\caption{Pavement digital surface acquisition model.}\label{fig:acquisition}
\end{figure}

An example of data acquisition by CROP method is given in Fig.~\ref{fig:acquisition}. While the camera arrangement is shown in Fig.~\ref{fig:crop}, the created pavement digital surface models (DSMs) can be observed in Fig.~\ref{fig:surface_crop} and~\ref{fig:surface_digital}. Once the DSMs were created, they were subjected to further processing in the open-source software for dense point cloud analysis Cloud Compare, v2.11.3. The DSMs of investigated surfaces were subjected to initial adjustments: leveling, so the 3D models were parallel to the horizontal plane; scaling, to correspond to millimeters unit; and filtering, so the DPC points falling outside the area of interest or relevant texture scale were removed. The final DPC model of each surface was subjected to profile segmentation to investigate the roughness properties of analyzed surfaces.


\begin{figure}[b]
    \centering
    \includegraphics[width=0.6\linewidth]{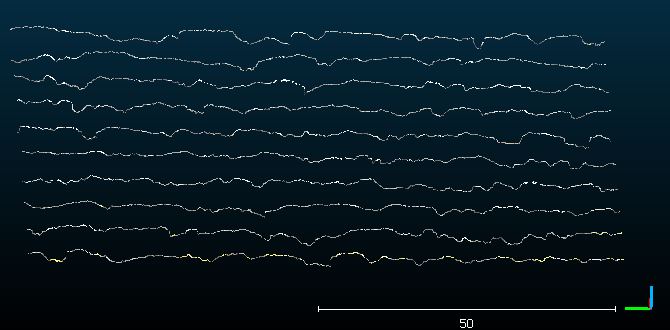}
    \caption{Profiles segmented from the digital surface model with defined segmentation properties: profile length is $100\,\si{\milli\meter}$ and profiles' lateral distance is $10\,\si{\milli\meter}$.}
    \label{fig:Fig3c}
\end{figure}

To determine the roughness properties of pavement surfaces that could be compared to the friction performance evaluated by the proposed numerical method, an analysis of profile roughness features was conducted. The profiles were segmented from the DSMs of investigated surfaces in Cloud Compare software. To segment the profiles, it has been necessary to define a segmentation area, a segment orientation, a segment length (i.e., profile length), and a lateral distance and thickness of the segmented DPC sections. The segmentation area was equal to the pavement surface area inside the reference frame and the segments were oriented to be parallel to a horizontal $x$-axis. The length of the segment in the $x$-direction was set to be $100\,\si{\milli\meter}$, to correspond to the profile's length for the calculation of the traditional pavement texture indicator mean profile depth (MPD) following the EN ISO 13473-1 standard norm. The segments' lateral distance was set to $10\,\si{\milli\meter}$, so that each investigated surface was described by ten profiles, as depicted in Fig.~\ref{fig:Fig3c}.

\begin{figure}
    \centering
        \subfloat[][Profile segment with section thickness equal to $0.005\,\si{\milli\meter}$.\label{fig:Fig4a}]
            {\includegraphics[width=0.6\textwidth]{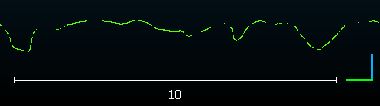}}\\
        \subfloat[][Profile segment with section thickness equal to $0.010\,\si{\milli\meter}$.\label{fig:Fig4b}]
            {\includegraphics[width=0.6\textwidth]{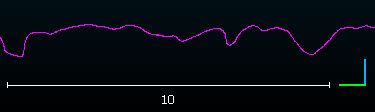}}\\
        \subfloat[][Profile segment with section thickness equal to $0.050\,\si{\milli\meter}$.\label{fig:Fig4c}]
            {\includegraphics[width=0.6\textwidth]{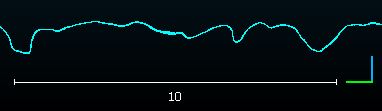}}
    \caption{The same $10\,\si{\milli\meter}$ length profile segmented with three different section thicknesses.}\label{fig:section_thickness}
\end{figure}

The depth in the direction orthogonal to the average surface mean plane of the segmented section was set to be $0.01\,\si{\milli\meter}$, selected after the comparison of profile's number of points for different section depth values settings. This value determines the width of the DPC segment from which the profile is created. If the selected depth values were too wide, the profile would consist of points having a single $x$ coordinate and multiple $z$ coordinates corresponding to the same $x$ value. This would require dataset interpolation to obtain a 2D profile and such approximation could influence the true geometry of the surface. On the other hand, for a too narrow section depth, the number of points in a profile would not be sufficient to properly describe all roughness features as such profiles would have gaps, requiring, again, extrapolation and true roughness features approximation. An example of a profile segmented with three different section depths is given in Fig.~\ref{fig:section_thickness}, obtained from the Cloud Compare software as result of the DPC data analysis.

For a selected section depth of $0.01\,\si{\milli\meter}$, the obtained profiles showed to have an average point-to-point distance in horizontal direction below $0.01\,\si{\milli\meter}$. This value implies that extracted profiles have sub-millimeter resolution in horizontal direction, which enables the roughness features analysis on both micro- and macro- texture scales relevant for the analysis of pavement friction phenomenon. The profiles extracted with the previously described settings in the segmentation procedure had roughly $12.000$ points on average. The exact number of points in each profile differs due to the morphology varieties of inspected surfaces in the positions where the profiles were segmented. For the purpose of roughness features comparison of extracted profiles, all profile points coordinates were converted from a relative to an absolute reference coordinate system so that each profile starts with a point having $x$ coordinate equal to zero, while height coordinates were adjusted so that the minimum height value was set to zero. 
After the profiles' coordinates were corrected, the adjusted profiles were imported to Mountains Map, v9.0 Lab Premium software, for analysis of the roughness features. The software enables automatic calculation of profile's texture roughness features based on the EN ISO 21920-2 standard. Roughness features can be calculated on profiles with or without scale separation, a feature that enables the exclusion of specific texture scales, if desired. In this research, no scale separation was applied and the texture parameters were determined on primary profiles containing both the relevant texture scales. To remove any remaining profile vertical slope, the profiles were automatically leveled to a horizontal plane. A Gaussian S-filter was applied to the leveled profiles to remove the elements in lateral direction which were not below a threshold of $2.5\,\si{\micro\meter}$. In this way, the profile dataset was de-noised, while the relevant profile points within the defined horizontal resolution of $0.01\,\si{\milli\meter}$ were preserved.
Profile related roughness parameters were selected from the group of \emph{amplitude}- or \emph{height}-related parameters and \emph{feature} parameters group, defined in the EN ISO 21920-2 standard. They are the most common non-standard roughness parameters groups explored in pavement texture analysis in relation to frictional performance~\cite{bitelli:2012, callai:2022, zuniga:2019, kogbara:2018}. The amplitude parameters are a sub-group of field parameters, related to the full length profiles and calculated from the profile heights $z(x)$. Feature parameters are calculated for a given section of a full length profile, describing amplitude or wavelength features of profiles' sections within specific section lengths. Additionally, the traditional pavement texture profile-related parameter MPD was calculated for the analyzed profiles to compare its values to the values of the obtained non-standard parameters. Table~\ref{tab:Table1} provides an overview of the roughness parameters calculated for the analyzed profiles, with a description of their physical meaning.

\begin{table}
    \centering
    \caption{Profile roughness parameters determined by following EN ISO 21920-2 standard.}
    \label{tab:Table1}
        \begin{tabular}{p{30mm} p{40mm} p{45mm}} 
            \midrule
            Parameter ($\si{\milli\meter})$&Group&Description \\ 
            \midrule
            $\Pa$ & amplitude & arithmetic mean height for full profile length\\  
            $\Pq$ & amplitude & root mean square height for full profile length\\  
            $\Pt$ & amplitude & total height difference for full profile length\\ 
            \midrule
            $\Ppt$ & feature (profile peak) & maximum peak height for all profile sections\\  
            $\Pvt$ & feature (profile peak) & maximum pit depth for all profile sections\\  
            $\Pz$ & feature (profile peak) & mean value of maximum total height difference on all profile sections\\
            \midrule
            $\Psm$ & feature (profile element) & mean profile element spacing in horizontal direction for all profile elements\\  
            $\Psmx$ & feature (profile element) & maximum profile element spacing in horizontal direction for full profile length\\  
            $\Pc$ & feature (profile element) & mean profile element height for all profile elements\\  
            $\Pcx$ & feature (profile element) & maximum profile element height for all profile elements\\  
            \midrule
            $MPD$ & traditional parameter & mean profile depth from two highest profile peaks in the first and second half of the full profile length\\ 
            \bottomrule
        \end{tabular}   
\end{table}

\subsection{Selection of profiles for numerical simulation}\label{sec:profile_sample}

The aim of the research is to assess if texture roughness features calculated from the processed pavement surface profiles relate to the friction response obtained for the same profiles, when they are utilized as input sliding support in the numerical simulations. For this purpose, three profiles were selected arbitrarily from the database with different values of calculated texture parameters, extracted from three pavement surface models obtained by the CROP method. The surfaces were previously chosen for their friction performance after a standard pavement friction measurement campaign performed with the aid of a static skid resistance tester (SRT), test~\emph{EN ISO 13036-4}. This device is used for the measurement of low-speed friction performance of pavements based on the pendulum principle~\cite{hall:2009}. The obtained friction values are expressed as SRT, a dimensionless number ranging from $0$ to $150$, where higher values indicate better friction performance of the inspected surface. Three surfaces selected from the database showed different measured values of SRT: surface P06 exhibited SRT of $75.2$, surface P12 SRT was $85.4$ and, for surface P20, the evaluated SRT was $94.4$. To be able to compare the roughness properties of the surfaces evaluated by the SRT device, the CROP method was later applied on the exact same surface area. The size of the inspected surface was $125\,\si{\milli\meter}\times 75\,\si{\milli\meter}$, corresponding to the sliding distance of the pendulum's skid and the skid's width.
The analyzed profiles were selected from the central part of the surface model, as the CROP method verification procedure showed that the best precision of a DSM is obtained in the center of the model. All profiles were extracted from the DSMs by following the same procedure as described in the previous section, with resulting profile properties: profile length $100\,\si{\milli\meter}$, profile resolution $0.01\,\si{\milli\meter}$, obtained with a low-pass filter of $2.5\,\si{\micro\meter}$. 

\begin{figure}
    \centering
    \includegraphics[width=\textwidth]{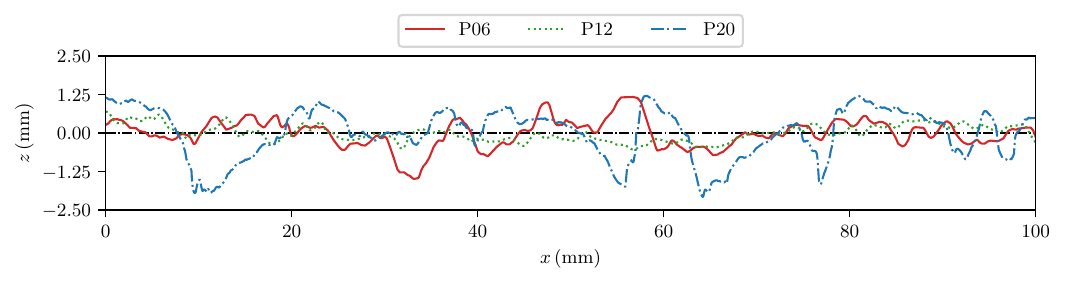}
    \caption{Profiles elevation fields.}
    \label{fig:raw_profiles}
\end{figure}

The geometry of selected profiles is provided in Fig.~\ref{fig:raw_profiles} and the resulting roughness parameters values on the inspected primary profiles, calculated following the relevant standards, are given in Tab.~\ref{tab:Table2}. From Tab.~\ref{tab:Table2} it can also be seen that profile P12 obtained the lowest values of roughness parameters in most of the amplitude and feature parameters group. The exception are values for $\Psm$ and $\Psmx$ feature parameters, which are related to the profile's morphology in horizontal direction. The traditional texture parameter $MPD$ is the lowest for P12 profile and highest for P20 profile.

\begin{table}
    \centering
    \caption{Measured friction values expressed in SRT $(-)$ and calculated profile roughness parameter values for three analyzed profiles; all roughness parameters values are expressed in $(\si{\milli\meter})$.}\label{tab:Table2}
    \vspace{2mm}
    \begin{tabular}{lrrr} 
        \toprule
        Profile & P06 & P12 & P20 \\
        \midrule
        $SRT$ & 75.2& 85.4& 94.4\\
        $MPD$ & 1.091& 0.563& 1.184\\
        $\Pa$ & 0.352& 0.211& 0.663\\
        $\Pq$ & 0.459& 0.258& 0.809\\
        $\Pt$ & 2.665& 1.233& 3.285\\
        $\Ppt$ & 1.180& 0.702& 1.214\\
        $\Pvt$ & 1.486& 0.531& 2.070\\
        $\Pz$ & 1.374& 0.724& 2.393\\
        $\Psm$ & 12.319& 12.586& 11.802\\
        $\Psmx$ & 27.522& 40.029& 18.518\\
        $\Pc$ & 1.120& 0.565& 2.100\\
        $\Pcx$ & 2.094& 0.844& 3.285\\
    \bottomrule
    \end{tabular}
\end{table}

Profile P20 obtained the highest values for all calculated parameters, in coincidence with the highest $SRT$ value determined for the surface from which this profile was segmented. However, the measured friction values are surface-related and therefore cannot be a reliable representation of profile's roughness characteristics analyzed in this work. Amplitude parameters $\Pa$ and $\Pq$ calculated as mean height values for the full profile length showed less variation in values among profiles in comparison to feature parameters describing the peak profile features, such as $\Pz$ or $\Pvt$. The highest variation of parameters values was obtained for $\Pcx$ profile element parameter, characterizing the maximum profile element height. 

Considering the obtained values of roughness parameters calculated for the three segmented profiles, it can be concluded that the profiles have significantly different morphology. Consequently, it is expected for them to show different friction performance in the following numerical simulations.

\section{Computational framework}\label{sec:comp_fram}
The computational framework chosen to assess correspondences between sample roughness parameters and friction arising between asphalt and tire leverages on a novel interface finite element introduced in~\cite{paggi:2018} and called MPJR, eMbedded Profile for Joint Roughness. The method hinges on a special class of finite elements originally derived for cohesive zone models applied in the field of non-linear fracture mechanics. It allows, in its basic formulation, to easily analyze contact problems involving rough entities coming into contact. In subsequent extensions, the method has been further developed to account for infinitesimal sliding with Coulomb friction~\cite{bonari:2021} in two and three dimensions, for both non adhesive and adhesive contact problems~\cite{bonari:2022}, with the extended possibility of considering any arbitrary geometry for the rough surface, whose height field could be defined according to analytic functions or synthetic or experimental data. Furthermore, the context of finite sliding has been investigated as well in~\cite{bonari:2020}, where a seminal approach has been tested to simulate finite sliding of a linear viscoelastic block over a rigid surface with a simple harmonic geometry.

In this study, the last two show-cased features of finite sliding of a linear viscoelastic body on real rough profiles, i.e., whose geometry is defined through an experimental campaign of data acquisition, are combined together for the first time, resulting in a step further in the field of high fidelity simulations of tire-asphalt interaction. The phenomenon is governed, at the micro-scale, by the interplay of different friction mechanisms. On one hand there is \emph{Coulomb} or \emph{dry friction}, which is proportional to the normal force that brings the surfaces in contact and is characterized by a coefficient of friction that remains relatively constant regardless of the contact area and the sliding velocity~\cite{bowden:1950}. In contrast, \emph{rubber friction} involves more complex interactions, due to rubber's viscoelastic properties. This type of friction is highly dependent on the texture of the contacting surface, on the applied pressure, on the sliding velocity and on the temperature~\cite{persson:1997}. Rubber friction itself consists, in turn, of two main components: adhesion, since rubber tends to adhere to the contacting surface, and hysteresis, an energy loss due to the inelastic deformation of the polymeric chain composing the microstructure of the rubber that results in heat dissipation. As a result, rubber friction can vary significantly with changes in these factors, making it less predictable and more complex than Coulomb friction.

Since the proposed computational approach already proved reliable in the analysis of both Coulomb friction and adhesive problems, the main scope of the current investigation is devoted to the analysis of frictional effects due to hysteresis alone. A positive outcome of the current investigation will then pave the way for the derivation of a comprehensive interface model capable of bringing together all the aforementioned features, namely, Coulomb friction and rubber friction due to both adhesion and hysteresis.

To the scope, a simulation framework has been set up, where the tire has been modeled as a linear viscoelastic block pressed against a rigid rough profile and moved under the action of an imposed horizontal act of motion, characterized by specific values of velocity. The numerical solution is obtained in terms of the hysteretic friction coefficient for different sliding velocities employing, as sliding support, the different profiles obtained in Sec.~\ref{sec:roughness}.

\subsection{Rheological model for rubber block}\label{subsec:material_model}
Real viscoelastic materials exhibit a transient mechanical response that varies significantly across multiple orders of magnitude of time and intensity. As a result, a simplistic model comprising a single linear Hookean spring in conjunction with a Newtonian dashpot fails to accurately represent their behavior. For engineering applications, typically at least three terms in the Prony series are required to yield meaningful stress analysis predictions. In our case, the classic expression for the Young's relaxation modulus has ben employed:

\begin{equation}\label{eq:rheo}
E(t) = E_0+\sum_{n=1}^3 E_n\exp{\Bigl(-\frac{t}{\tau_n}\Bigl)}.
\end{equation}

The rheological properties of the material can be investigated in the frequency domain performing a Fourier transform of the relaxation modulus defined in Eq.~\eqref{eq:rheo}. The complex modulus in the frequency domain reads:

\begin{equation}
    \hat{E}(\omega) = E_0 + \sum_{i=1}^n E_i\frac{\tau_i^2\omega^2}{1+\tau_i^2\omega^2} + \imath \sum_{i=1}^n E_i\frac{\tau_i\omega}{1+\tau_i^2\omega^2}.
\end{equation}

Starting from the complex modulus, the storage modulus can be defined as the real part $\Re{\hat{E}}(\omega)$ of Eq.~\eqref{eq:rheo}, and it describes the ability of the material to store elastic energy. Specularly, the loss modulus is defined as the imaginary part $\Im{\hat{E}}(\omega)$ of $\hat{E}(\omega)$ and describes the part of energy which is dissipated as heat as the material is subjected to deformation.



\begin{figure}
    \centering
    \subfloat[][Response modulus for three, solid red line, and six terms, circular markers, of the Prony series.\label{fig:first_prony}]
        {\includegraphics[width=0.45\textwidth]{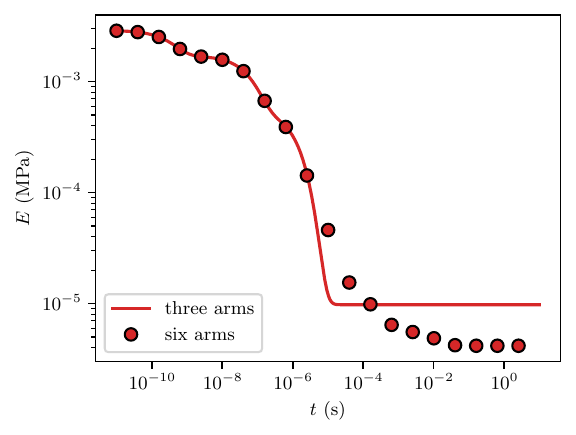}}
    \\
    \subfloat[][Storage modulus for the three arms model, solid curve, the six arms model, circular markers, and for experimental data, dotted line.\label{fig:second_moduli}]
        {\includegraphics[width=0.47\textwidth]{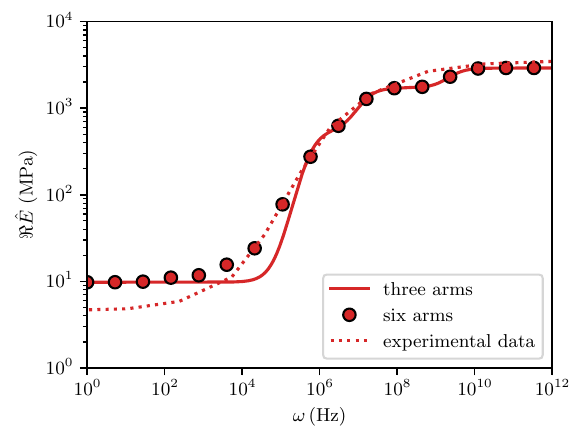}}
    \hspace{5mm}
    \subfloat[][Loss modulus for the three arms model, solid curve, the six arms model, circular markers, and for experimental data, dotted line.\label{fig:third_moduli}]
        {\includegraphics[width=0.47\textwidth]{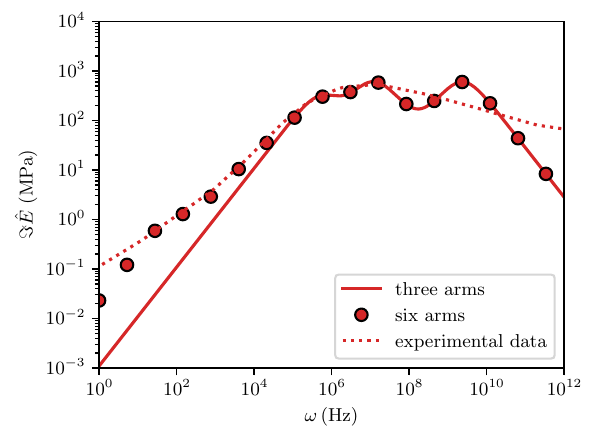}}
    \caption{Rheological diagrams for rubber like material.}
    \label{fig:complex_rheology}
\end{figure}

Due to the limitation imposed by the FEM solver in terms of material parameters definition, a model characterized by a Prony series with three arms has been chosen. To obtain a realistic response, the material model employed both in~\cite{wriggers:2009} and~\cite{delorenzis:2013}, which in turns is based on real data of Styrene Butadiene Rubber (SBR) acquired during an experimental campaign performed at the German Institute for Rubber Technology, has been fitted using Eq.~\eqref{eq:rheo}.
A comparison between our model and the model employed in the two aforementioned studies is depicted in Fig.~\ref{fig:complex_rheology}. More specifically, Fig.~\ref{fig:first_prony} shows a comparison between two different realization of the relaxation modulus characterized by three arms, continuous red line, and six arms, red circular markers. On the other hand, Figs.~\ref{fig:second_moduli} and~\ref{fig:third_moduli} shows a comparison between averaged real data, a three arms model, continuous lines, and a six arms model, circular markers. In all the plots, green color refer to the loss modulus, while red color to the storage modulus. It can be noticed how the models characterized by a different numbers of the terms composing the Prony series deliver identical results at higher frequency responses, while differing at low excitation frequencies.

\subsection{Friction due to hysteresis}
Let us assume a linear viscoelastic body pressed against a rough profile and sliding over it under the action of a tangential load. If both adhesive and Coulomb frictional effects are neglected, the contact tractions will always be orthogonal to the deformed surface. In the framework of a linear theory, a global time varying horizontal contact force $Q(t)$ can then be determined via a geometric relation, as the integral of the projection of the normal contact tractions $\pn$ over the surface gradient $\up$. If $P(t)$ is the overall time varying vertical reaction force, a global averaged coefficient of friction can then be defined as:

\begin{equation}\label{eq:mu}
\omu = \frac{1}{T}\int_0^T\frac{Q(t)}{P(t)}\,\mathrm{d}t 
               = \frac{1}{T}\int_0^T\frac{\int_\Gamma \pn \up \,\mathrm{d}x}
                                         {\int_\Gamma \pn \,\mathrm{d}x}\,\mathrm{d}t.
\end{equation}

For problems characterized by simple geometrical and mechanical features, an analytical expression can be derived for Eq.~\eqref{eq:mu}.
For example, let us consider a layer of viscoelastic material, characterized by a single relaxation time $\tau$, that extends indefinitely in the horizontal direction and has a finite depth $l$. If it makes full contact with a sinusoidal profile of wavelength $\lambda$ and amplitude $a$, under the action of an imposed vertical far-field displacement $u_0$, and slides over it with constant velocity $v$, the average coefficient of friction can be derived as:

\begin{equation}\label{eq:sine}
    \omu = \frac{\pi E_1 a^2 \omega \tau}{E^\infty u_0 \lambda (1+\omega^2\tau^2)},
\end{equation}

where $\omega = 2\pi v/\lambda$ is the excitation frequency and $E_1$ and $E_\infty = \sum_i E_i$ are the long term and the instantaneous elastic modulus of the material, respectively. From this preliminary simple model, an important property of the viscoelastic coefficient of friction can be deduced, i.e., its marked dependency on the sliding velocity $v$. For this simple case it is straightforward to conclude that the maximum value of $\omu$ is reached when the system is excited with a frequency $\omega^\star = 1/\tau$. In a more general scenario, characterized by additional complexities such as more complex material laws, partial contact or more complex support sliding surfaces, no closed form expression can be derived and the mean value of the hysteretic friction coefficient must be evaluated numerically according to Eq.~\eqref{eq:mu}.

\begin{table}
    \centering
    \caption{Mechanical parameters for the three arms Maxwell Model.}
    \vspace{2mm}
    \begin{tabular}{lr}
        \toprule
        $E\,(\si{\mega\pascal})$ & $\tau\,(\si{\second})$ \\
        \midrule
        $9.77$ & $-$ \\
        $5.41\times10^02$ & $1.85\times10^{-06}$ \\
        $1.16\times10^03$ & $8.09\times10^{-08}$ \\
        $1.19\times10^03$ & $4.22\times10^{-10}$ \\
        \bottomrule
    \end{tabular}
    \label{tab:mech_parameters}
\end{table}

\subsection{MPJR interface finite element for normal contact}
In its 2D formulation, the MPJR interface finite element is a four nodes quadrilateral element originally defined to model interfaces in the context of nonlinear fracture mechanics, addressed using a cohesive zone model~\cite{paggi:2011a,paggi:2011b} and later on extended to contact mechanics~\cite{paggi:2018}. This framework allows to solve contact problems involving deformable rough surfaces in contact in a straightforward fashion, avoiding to explicitly model the features of geometrically complex entities, exploiting the concepts of composite topography and composite mechanical parameters~\cite[\S2.2.3]{barber:2018}. Within a small deformation setting, the normal contact of two elastic bodies characterized by a complex interface can be reformulated as a contact problem involving a rigid rough body, whose shape is a combination of the two original geometries, and a linear elastic body with a flat interface whose mechanical parameters are a combination of the original ones. If the original bodies are already rough-rigid and flat-deformable, this reformulation step is unnecessary. Since this condition is coincident with the current case, and no composite mechanical parameters are involved, any material law can be defined for the deformable bodies, and in our case a linear viscoelastic material is employed to model the rubber rheology.

Since one of the body is rigid, it has not to be explicitly modeled and the domain definition is limited to the deformable body and the contact interface, whereby the former can be discretized employing standard finite elements and the latter a single array of four nodes MPJR interface finite element. In each element, the two lower nodes are tied to the discretization of the deformable body, while the two upper nodes belong to the rough profile. 

\begin{figure}[b]
    \centering
    \includegraphics[width=0.3\textwidth]{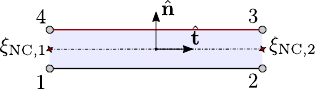}
    \caption{Kinematics of the 2D quadrilateral MPJR element.}
    \label{fig:2d_mpjr}
\end{figure}

The kinematics of the element is shown in Fig.~\ref{fig:2d_mpjr}. An array of unknown horizontal and vertical nodal displacements $\mathbf{u} = [u_1,v_1,\dots,u_4,v_4]^\intercal$ is introduced for the evaluation of the gap function $\mathbf{g} = [\gtg,\,\gn]^\intercal$ across the interface, that reads:

\begin{equation}
    \mathbf{g} = \mathbf{Q}\mathbf{N}\mathbf{L}\mathbf{u},
\end{equation}

where $\mathbf{L}$ is a linear operator that computes the relative nodal displacements across the interface in horizontal and vertical direction, $\mathbf{N}$ is a matrix that collects standard first order shape functions and $\mathbf{Q} = [\hat{\mathbf{t}},\,\hat{\mathbf{n}}]^\intercal$ is a rotation matrix that maps the gap function from a local reference system, aligned with the element, to a global $\Omx$ reference frame, where $\hat{\mathbf{t}}$ and $\hat{\mathbf{n}}$ are the unit tangential and normal vectors aligned with the element centerline and still evaluated in $\Omx$.

According to this formulation, the element is suitable for the solution of contact problems involving smooth conformal interfaces. At this point, a correction in the nodal 
vector $\mathbf{u}$ is performed by hard coding the position of the two upper nodes $v_3^\star$ and $v_4^\star$ to match the actual position of the rigid rough surface considered. The result of this operation is a corrected gap function whose normal component $\gn^\star$ describes the kinematics of the real geometrically complex problem, now reading:

\begin{equation}
    \gn^\star = \hat{\mathbf{n}}\cdot
    \begin{bmatrix}
        N_1 (u_4-u_1) + N_2 (u_3-u_2) \\
        N_1 (v_4^\star-v_1) + N_2 (v_3^\star-v_2)
    \end{bmatrix}
    \label{eq:qnlu}
\end{equation}

Finally, the externally applied contact tractions can be evaluated, e.g., with a standard penalty approach governed by a penalty parameter $\en$:

\begin{align}
    \pn = 
    \begin{cases}
        \en\gn^\star \quad &\text{if} \quad \gn^\star <   0  \\
        0            \quad &\text{if} \quad \gn^\star \ge 0
    \end{cases}
\end{align}

A sketch of the whole process is qualitatively shown in Fig.~\ref{fig:mpjr_method} below. It has to be remarked that in the context under examination the degrees of freedom $u_i$ related to horizontal displacements do not play any active role in the simulation, since the layer of the interface finite elements is disposed on a horizontal line according to the global reference frame, so that no related term appears in Eq.~\eqref{eq:qnlu}, and no Coulomb friction is considered acting at the interface level. For a complete derivation of the method, together with validation and quality assessment of the results, the interested reader is referred to~\cite{bonari:2021}.

\begin{figure}[b]
    \centering
    \subfloat[][Original problem setting.\label{fig:method_cont}]
        {\includegraphics[width=0.45\textwidth]{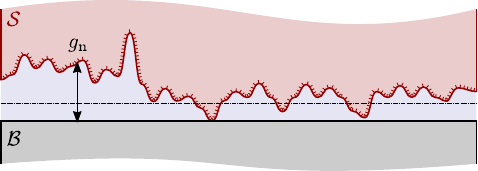}}
    \hspace{5mm}
    \subfloat[][Discretized setting.\label{fig:method_disc}]
        {\includegraphics[width=0.45\textwidth]{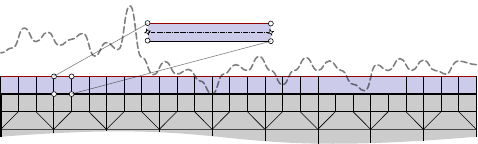}}
    \caption{Sketch of the MPJR approach.}
    \label{fig:mpjr_method}
\end{figure}

\subsection{MPJR interface finite element for predicting finite sliding}
In the current study, the introduced MPJR approach is extended with an additional feature that allows to consider a relative finite sliding involving a rough profile expressed as a set of discrete elevations over a viscoelastic body. Starting from the derivation given in~\cite{bonari:2020}, this is possible thanks to the main hypotheses on which the formulation hinges, which are:

\begin{enumerate}
    \item rigidity of the rough profile;
    \item apriori knowledge of the profile's act of motion;
    \item small displacements analysis.
\end{enumerate}

Given these assumptions, it is always possible to express, for every analysis time step, the relative position of the viscoelastic body and the sliding profile, and therefore the relative position of the two~\emph{leading} and~\emph{trailing} closest points of the profile with respect to the correspondent nodes of the boundary of the deformable body, i.e., the two lower nodes of each interface finite element.

The sliding viscoelastic body, hereinafter labeled as~\emph{skid}, is modeled as a rectangular domain, in the reference frame $\Omx$ set in correspondence of the skid's top-left corner and with the $x_1$ axis aligned to its top side. The analysis is developed over a time window $T=[t_0,\,\tf]$ and is divided in two main phases, a~\emph{Phase I} involving vertical loading, in which the rough profile is pressed against the skid, and a~\emph{Phase II}, where a horizontal velocity is imposed, which determines the sliding motion. Section~\ref{sec:loading_process} offers a detailed description of the two different phases of the loading process.

Sliding is supposed to take place between the rough profile and the top side of the skid, with the rough profile moving in accordance with a predefined rigid translating act of motion $\yo$ characterized by a given horizontal and vertical velocity. As stated in Sec.~\ref{sec:profile_sample}, every profile geometry is stored as a set of discrete points $\yp$, whose first coordinate is set so that, considering a local reference frame $\Oy$ solidal with the rough profile, the first point of the ensemble is characterized by $y_\mathrm{p1}=0.0$. At $t=0$, the rough profile makes contact with the skid at its minimum point only, without exerting any contact pressure. In the skid's global reference frame, the position of each point of the rough profile can be then expressed as $\xp$.

\begin{figure}[t]
    \centering
    \includegraphics[width=0.5\textwidth]{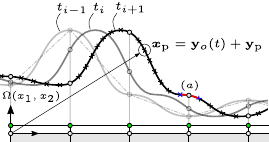}
    \caption{Correct nodes elevation according to their actual position during the sliding process. The deformable bulk is shaded in gray, while the array of MPJR elements is highlighted on its top by its four nodes. For each element, the two top nodes, highlighted in green, do not enter the calculations, while the actual top nodes' position is inferred at each time step. Cross markers indicate the sampled points on the rough profile, and the nodal elevation used to correct the normal gap is evaluated from a spline interpolation between the leading and the trailing points next to each node. In the figure, this operation is highlighted for node $(a)$. The leading and trailing points are highlighted in blue, while the interpolated portion of the profile is depicted in red. The sliding process is visualized by the superposition of different profiles snapshots in different shades of gray, each referred to a single time instant, the lightest being related to the time step furthest in time.}
    \label{fig:skid_crs}
\end{figure}

To reproduce the right contact conditions, an array of interface finite elements is defined over the skid's top side, where, in accordance with the MPJR paradigm, the lower pair of nodes is tied to the FEM discretization of the skid's boundary, while the normal gap $\gn$ is corrected to account for the variation of the position of the rough profile, Fig.~\ref{fig:skid_crs}.

While for a profile defined by an analytical expression the correction of the normal gap can be performed straightforwardly via a parametrization in time of the function describing the profile's elevation, cfr. Eq.~$(12)$ in~\cite{bonari:2020}, the same operation can not be extended to the current case, since the elevation is known only at discrete points, thus allowing to correct the normal gap only if the horizontal position of a profile's point is coincident with the horizontal position of the skid's boundary node. To overcome the problem, a cubic spline interpolation is performed once in a first offline stage, to sample the intervals between $y_{\mathrm{p}1,i}$ and $y_{\mathrm{p}1,i+1}$ and, for every point $i$, the interpolation coefficients $c_{i,j} \,\text{for}\,j=[1,\dots,4]$ are stored beside the profile's horizontal position and elevation.

For every interface finite element considered, two elevation values have to be identified, one for each of the base nodes of the interface finite element. The correct elevation value is then inferred identifying the closest~\emph{trailing} and~\emph{leading} correspondent points from the profile's set. Once they have been identified, the correct elevation value is retrieved evaluating the interpolation in correspondence of the horizontal position of the interface finite element's node.

To keep the profile's horizontal velocity as generic as possible, the identification of the neighboring points is performed at each time step, but since the elevations are stored as an ordered list, a convenient binary search algorithm can be enforced to solve the problem. For each node, this operation must be carried out only once, since when the trailing point has been identified, the leading point is the following one, and vice versa. Once the elevation of the rough profile in correspondence of the interface node under examination has been carried out, its value is plugged into the expression of the corrected normal gap, cfr. Eq.~\eqref{eq:qnlu} and Eq.\,$(2)$ in in~\cite{bonari:2020}, and the resulting corrected gap can be employed to restore the true contact condition of the original problem.

\section{Numerical experiments setup}\label{sec:exp}

In this section, a first set of numerical experiments is performed to analyze the sliding of a linear viscoelastic block over a simple rigid sinusoidal shape. The profile is characterized by a wavelength $\lambda = 2\pi/320~\si{\milli\metre}$ and amplitude $g_0 = 2.0\times10^{-3}~\si{\milli\metre}$. The results in terms of $\mu(t)$ and $\omu$ are then compared with Eq.~\eqref{eq:sine} and the results presented in~\cite{delorenzis:2013}. Each feature of the model setup is then discussed in detail. Concurrently, these preliminary results are useful to tailor the model for the subsequent use of the rough profiles obtained during the campaign of experimental sampling.

\subsection{Rubber block geometry and FEM discretization}
\begin{figure}[b]
    \centering
    \subfloat[][Sketch of the problem under examination.\label{fig:methoda}]
        {\includegraphics[width=0.45\textwidth]{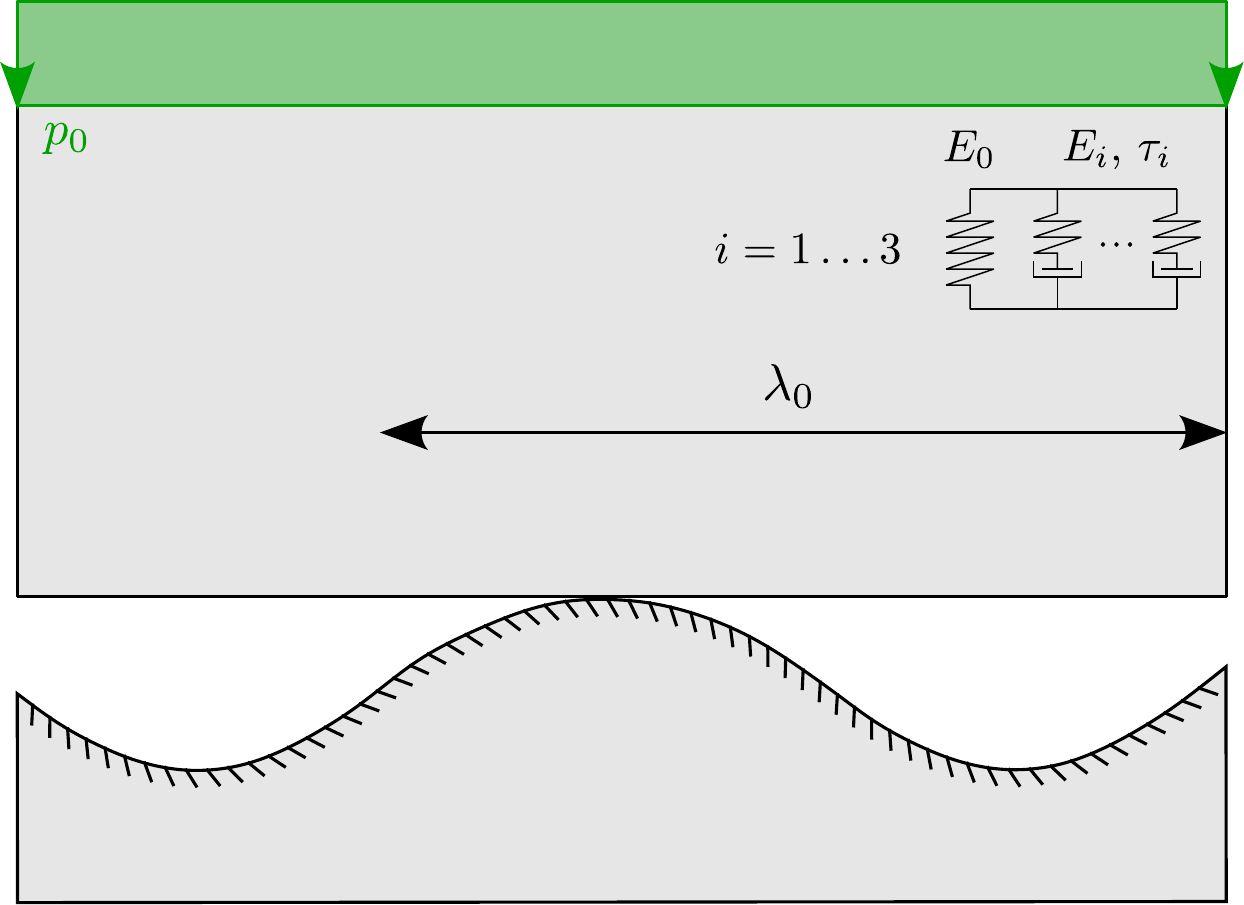}}
    \hspace{5mm}
    \subfloat[][Excerpt from numerical simulation at a selected representative time step.\label{fig:methodb}]
        {\includegraphics[width=0.45\textwidth]{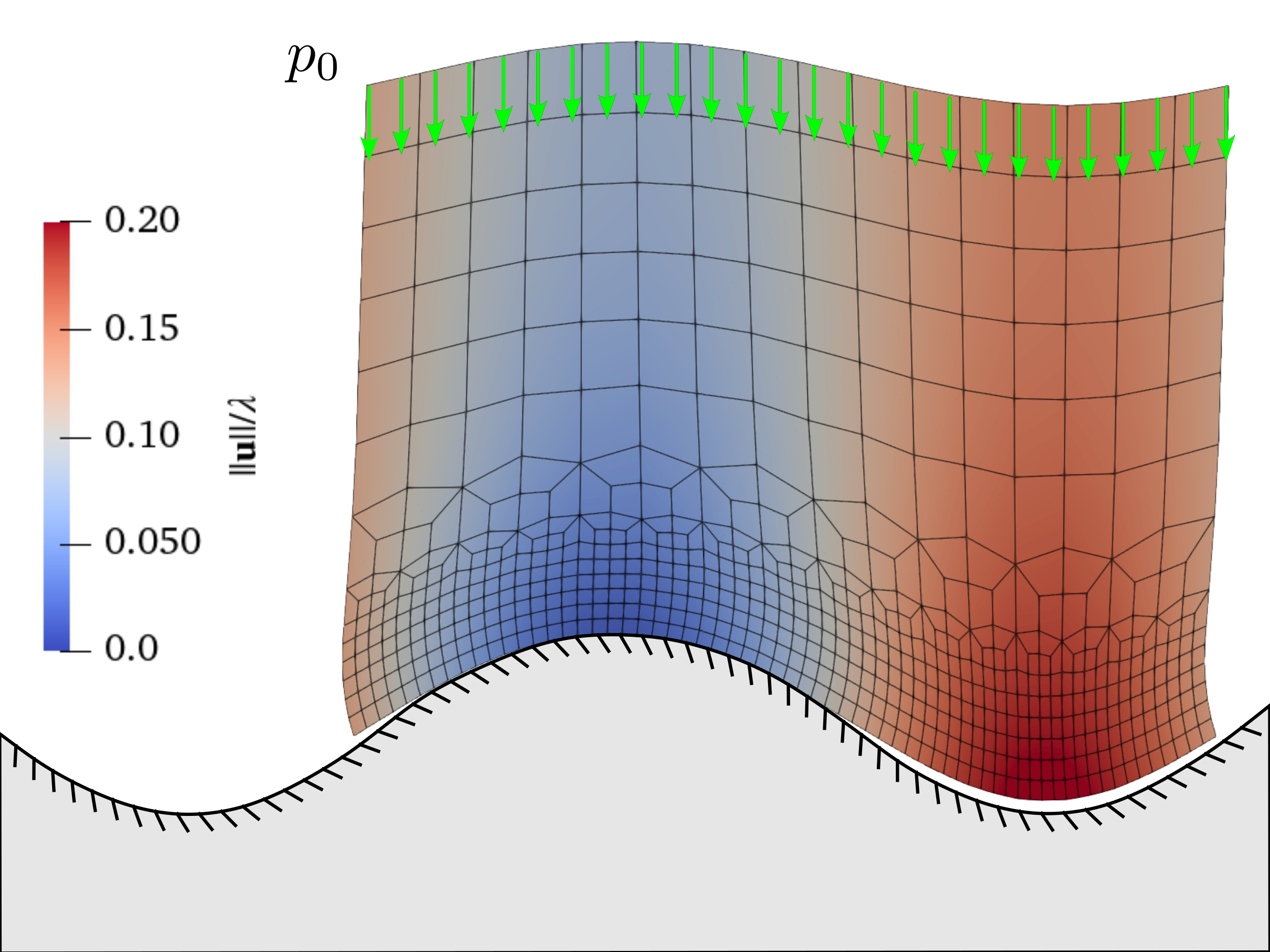}}
    \caption{Graphical representation of the addressed problem.}
    \label{fig:method}
\end{figure}

This part of the analysis focuses on a rectangular block with length $\bb=\lambda$ and $\hb = 0.75\lambda$. The domain is discretized employing standard linear quadrilateral elements, while an array of MPJR interface finite elements is placed in correspondence of the contact interface and stores the shape of the moving sinusoidal profile. To test the effect of an increasing refinement of the contact zone over the final value of the friction coefficient, the mesh is refined in correspondence of the contact side, where a different and progressively increasing number of elements has been employed. The results of the mesh convergence study are detailed in Sec.~\ref{sec:results_bench}, while a sketch of the problem under examination is depicted in Fig.~\ref{fig:methoda}, together with the typical mesh arrangement employed shown in an excerpt of a simulation run shown in Fig.~\ref{fig:methodb}.

The zone interested by mesh refinement, where bulk elements have the smallest size, is characterized by a certain depth. Below, the size of the mesh elements starts to increase hierarchically. Numerical experiments shows that this buffer zone must be characterized by a minimum depth value in order to achieve regular results in terms of contact tractions. By successive refinement, this minimum value has been qualitatively identified to be two times the size of the coarsest element in the mesh. In an analogous way, the penalty parameter $\en$ has been increased until the simulation results were not affected anymore by larger values, yet still avoiding divergence due to a too high $\en$. The final value $\en = 10^2E_\infty/\hb$ has been employed, being $E_\infty$ the long term Young's modulus of the viscoelastic bulk. 


\subsection{Loading process}\label{sec:loading_process}
The simulation is divided in two main stages. During the first, referred to as~\emph{Phase I}, the block is pressed against the rigid profile by a constant uniform pressure $p_0$ applied to the block on the opposite side with respect to the contact interface. In the following loading phase, namely~\emph{Phase II}, a uniform horizontal velocity $v$ is imposed on the top side, while $p_0$ is kept constant.

\subsubsection{Phase I}
\begin{figure}[b]
    \centering
        \subfloat[][Optimal values for the duration of the vertical loading phase for different values of excitation pulsation $\omega$.\label{fig:opt_t1}]
            {\includegraphics[width=0.45\textwidth]{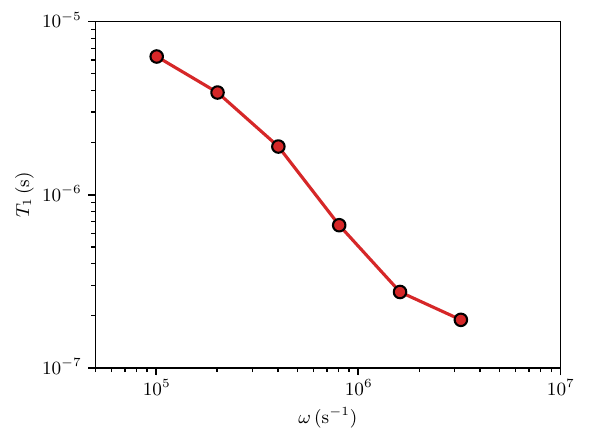}}
        \hspace{5mm}
        \subfloat[][Convergence of the average friction coefficient $\omu$ for different interface discretization levels and $\en=10E_\infty/h$.\label{fig:conv}]
            {\includegraphics[width=0.45\textwidth]{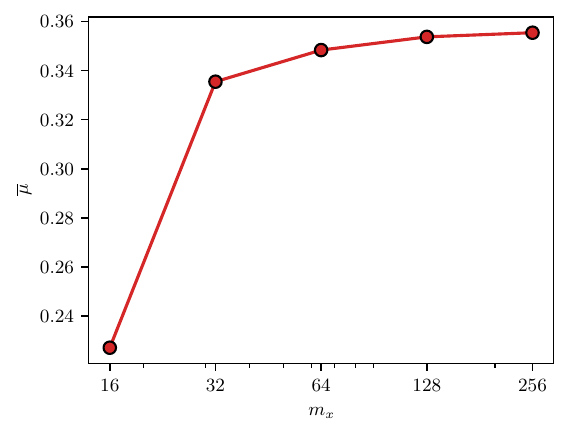}}
        \caption{Optimization in terms of vertical load velocity application and mesh refinement is performed to achieve reliable simulation results.}
    \label{fig:params}
\end{figure}

The most important parameter governing Phase~I is the velocity of application of $p_0$, quantified by the total duration $T_1$ of the vertical loading phase. It has been verified in~\cite{delorenzis:2013} that if the response of the system is more deformable during Phase~I than during Phase~II, then the system will need more time to reach a steady state during the sliding process. This will negatively affect the friction coefficient averaging process, since a wider analysis time window would be required. Therefore, a maximum value of $T_1$ has to be identified and, to this scope, the same approach used in~\cite{delorenzis:2013} has been employed. There, the desired value has been identified by solving the following optimization equation:

\begin{equation}
    \sum_{k=1}^nE_k\exp{(-T_1/\tau_k)} = 
    \sum_{k=1}^n E_k\frac{(\omega\tau_k)^2}{1+(\omega\tau_k)^2},
    \label{eq:opt_t1}
\end{equation}

where $T_1$ is obtained as a function of the applied velocity and $\bar{\omega}=(2\pi v)/\lambda$. The higher $v$, the lower must be $T_1$, i.e., the faster must the vertical load be applied. In case of simulations involving more sliding velocities, the more demanding condition has been addressed, coincident with the highest value of $v$. The related value $T_1$, i.e., the lower one resulting from the different velocities analyzed, has been used in every instance. After $T_1$ has been identified, a number of time steps $t_{\mathrm{s}1}$ sufficient to ensure the convergence of the solver has been set. Unless differently specified, a value of $t_{\mathrm{s}1} = 100$ is hereinafter used. The trend of the optimal value for $T_1$ in function of the block excitation frequency can be observed in Fig.~\ref{fig:opt_t1} for different values of $\omega$.

\subsubsection{Phase II}
At the end of Phase~I, the onset of sliding takes place and Phase~II begins. In every simulation, a constant value of sliding velocity is considered, while a transition phase with duration equal to $T_1$ is defined to guarantee a smooth acceleration of the sample obtained through a cubic fillet in the time velocity diagram. The duration of Phase~II is determined by the overall sliding length necessary to correctly evaluate $\omu$, a condition which is fulfilled by the full development of steady state sliding conditions. The adequate number of time steps to be considered depends on a proper sampling in time of the sliding profile. To ensure that, for each time step, the variation of the profile geometry is correctly captured, the number of time steps employed in the sliding phase is such that for each time increment, an advancement of the block on the profile of $0.2\bb/m_x$ takes place, where $m_x$ coincides with the number of interface finite element used to discretize the interface and $\bb/m_x$ is the distance between two consecutive points at which the height is sampled. If the total sliding distance is then expressed as a multiple $n_\lambda$ of the profile wavelength, the number of time steps necessary to cover the required distance results then in $t_{\mathrm{s}2} = 5 m_x n_\lambda$

\subsection{Results of the benchmark tests}\label{sec:results_bench}

In a first benchmark test, a sinusoidal profile has been employed. Two different conditions have been tested. In the first test case, a linear viscoelastic block with a single relaxation time has been employed. During Phase I, a uniform vertical load $p_0=10.0\,\si{\mega\pascal}$ has been applied in an overall time $T_1 = 0.16150688\,\si{\second}$, evaluated in accordance with the criterion expressed by Eq.~\eqref{eq:opt_t1}, and then horizontal sliding is enforced by the imposition of a constant horizontal velocity field $v$ to the profile.

In this very first preliminary test, a different number of interface finite element has been employed to discretize the interface and therefore the sinusoidal profile. Concurrently, different values of the penalty parameter have been tested, additionally. The results of the convergence study are presented in Fig.~\ref{fig:conv} for $m_x=16$, $32$, $64$, $128$ and $256$ interface finite elements employed. The penalty parameter $\en$ has been increased until the resulting traction field was not affected anymore by further increases, resulting in a value of $10E_\infty/h$, being $h$ the domain's height. In all the subsequent simulations, a value of $m_x=128$ interface elements per wavelength has been employed, since further refinements did not result in any appreciable accuracy gain in terms of $\omu$.

\begin{figure}[t]
    \centering
        \subfloat[][Instant value of the friction coefficient at critical sliding speed.\label{fig:bench_1}]
            {\includegraphics[width=0.45\textwidth]{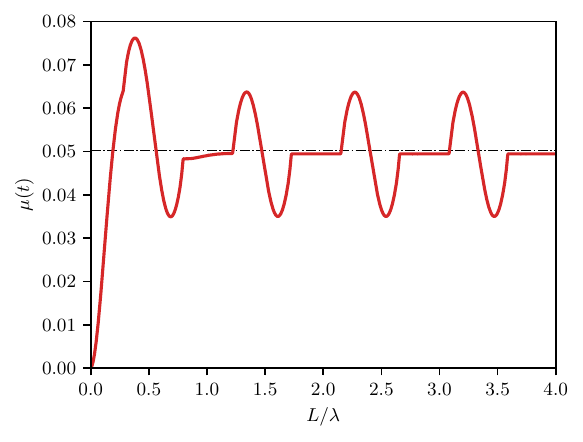}}
        \hspace{5mm}
        \subfloat[][Averaged values of the friction coefficient for different applied velocities.\label{fig:bench_2}]
            {\includegraphics[width=0.45\textwidth]{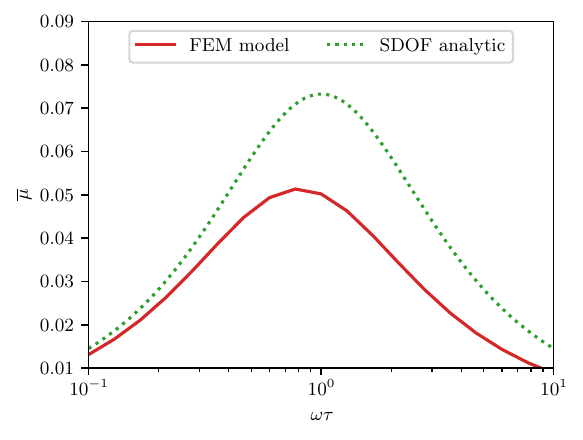}}
        \caption{Benchmark test, friction coefficient for different sliding velocities over a sliding distance $L=4\lambda$.}
    \label{fig:bench}
\end{figure}

The geometry of the block and the shape of the profile are the same listed at the beginning of Sec.~\ref{sec:loading_process}, and periodic boundary conditions are enforced in correspondence of the two vertical sides. The linear viscoelastic material employed for the test is characterized by a single relaxation time, with the following mechanical properties: $\nu=0.3$, $E_0 = 4.17\,\si{\mega\pascal}$, $E_1 = 1.72\,\si{\mega\pascal}$, $\tau_1 = 0.01134034\,\si{\second}$. Different values of the applied velocity have been tested, logarithmically centered around the critical value identified by $\omega^\star$, i.e., $v^\star = \lambda/(2\pi\tau_1)$. On the other hand, $T_1$ has been imposed to be correspondent to the most demanding condition, i.e., for the highest sliding velocity considered of $10\lambda/(2\pi\tau_1)$. The same value is then used in all the other cases, being them less demanding. Results are plotted in Fig.~\ref{fig:bench_1} in terms of $\mu(t)$, for the case $v=v^\star$. The results of the averaged value $\omu$ are plotted in Fig.~\ref{fig:bench_2} in function of every velocity considered, together with the values provided by Eq.~\eqref{eq:sine}. Good accordance is found, given the many different assumptions related to the two different models, above all: the analytical solution refers to a single degree of freedom model, while the numerical results to the discretization of a distributed setting.

\begin{figure}[b]
    \centering
    \includegraphics[width=0.45\textwidth]{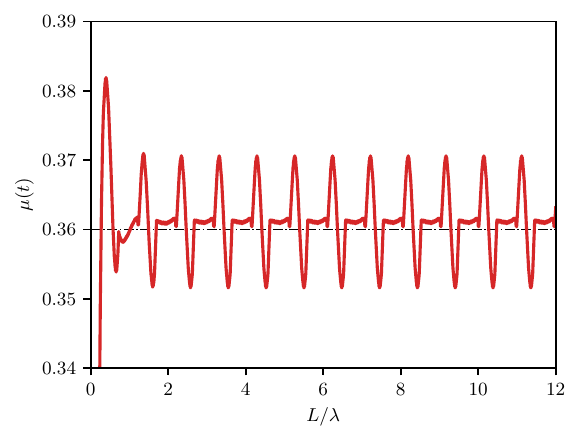}
    \caption{Sliding test for the imposed velocity causing a high dissipative effect, for an applied pressure of $p_0=10.0\,\si{\mega\pascal}$. The correspondent average value from~\cite{delorenzis:2013} is superposed as a dash-dotted line. The average value $\omu \approx 0.36$ corresponds to the result obtained in~\cite[\S 6.1]{delorenzis:2013} for the same values of applied pressure and imposed velocity.}
    \label{fig:benchmark_dl}
\end{figure}

In a second test, the same simulation setting has been employed, with a full viscoelastic model with three terms in the Prony series used to model the block's rheological properties. The material parameters are those identified in Sec.~\ref{subsec:material_model} and listed in Tab.~\ref{tab:mech_parameters}. Leaving the applied pressure unchanged, a velocity equal to $100\,\si{\milli\metre/\second}$, close to the critical value, is applied to the sliding profile. The related plot for $\mu(t)$ is shown in Fig.~\ref{fig:benchmark_dl}. A direct confrontation with the results from~\cite[\S 6.1]{delorenzis:2013} can be performed. The plot is compared with the related curve extracted from Fig.~10 (c) of the reference, for the case $h/\lambda = 0.75$. A perfect accordance can be found. It has to be remarked that such good accordance is to be expected only in correspondence of the highest viscoelastic response of the deformable material, since this is the region where the two different material models manifest the same rheological characteristics.

\section{Rough profile sliding}\label{sec:rough_sliding}
In this section, the hysteretic frictional properties of asphalt pavement surfaces are investigated using the extracted profiles as sliding support. Three different sets of simulations are performed, one set for each of the profiles analyzed, whose complete height field is depicted in Fig.~\ref{fig:raw_profiles}. Each simulation set is characterized by several test sliding velocities.

\begin{figure}[b]
\centering
    \subfloat[][Circular markers represent the original dataset alongside different level of down-sampling.\label{fig:conv_interp}]
    {\includegraphics[width=0.45\textwidth]{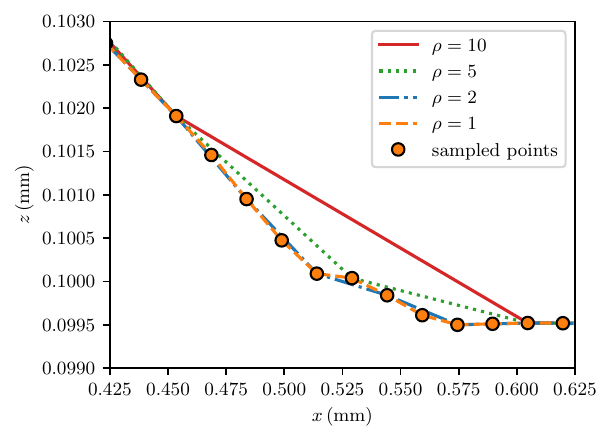}}
    \hspace{5mm}
    \subfloat[][Different realizations of $\mu(t)$ function for different sampling resolutions.\label{fig:conv_mu_rough}]
    {\includegraphics[width=0.45\textwidth]{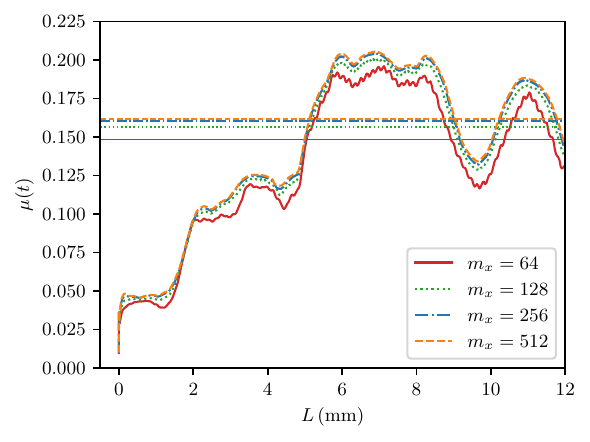}}
\caption{The down-sampling of the points acquired on the profiles reflects on the instantaneous value $\mu(t)$.}
\label{fig:conv_rough}
\end{figure}

\subsection{Profiles processing and resolution convergence study}

Given the high resolution of the sampling process, a direct correspondence between the acquired points and the interface finite elements storing the related height information rapidly leads to very high computational costs. For this reason, the effect of a down-sampling is analyzed first. For every profile, sub-profiles have been extracted by collecting one point every $\rho$ nodes, with $\rho = 1$, $2$, $5$, and $10$, respectively, where $\rho = 1$ corresponds to a condition for which every point of the profile is taken into consideration, thus leading to the highest achievable level of accuracy. The result of the down-sampling process can be observed in Fig.~\ref{fig:conv_interp} for a representative magnified portion of a sample profile and different chosen values of $\rho$. Since the skid discretization is, in turn, function of the number of points employed to describe the profile, the down-sampling process results in a mesh characterized by $m_x = 64$, $128$, $256$, and $512$ interface finite elements, respectively. Results in terms of $\mu(t)$ for the four different discretization levels are reported in Fig.~\ref{fig:conv_mu_rough}, for a simulation that considers a short test sliding length of $L=2b$ and for an applied velocity of $v=7\times10^3\,\si{\milli\meter/\second}$, while the other modeling assumptions and parameters are in line with the ones exposed in Sec.~\ref{sec:exp}. It can be noticed how a chosen value of $\rho=2$, implying the use of $m_x = 256$ interface finite elements, results in a value of $\omu$, dash dotted lines in Fig.~\ref{fig:conv_mu_rough}, very close to the one obtained with the finest grained resolution, which is obtained for $\rho = 1$. This chosen level of discretization will be employed in the final simulations related to the full scale models, i.e., for a sliding process that takes place over the whole length of the profiles.

\subsection{Duration of Phase I for rough profiles sliding}

Since the duration of Phase I is an important parameter to guarantee a rapid extinction of the transient period during the horizontal sliding, special care must be taken for its evaluation. For generic profiles a closed form expression analogous to Eq.~\eqref{eq:opt_t1} can not be obtained. If a rough profile is considered, it is not possible to follow the same strategy applied for the identification of an optimal $T_1$, as done for the sinusoidal profile. Nonetheless, a limiting value for the excitation frequency can still be identified if we consider the hypothetical profile with the shortest wavelength that can be taken into account given the discretization of the block's interface. If we consider this harmonic profile as a sampled time signal, then the shortest wavelength that can be sampled corresponds to its Nyquist frequency, i.e., two times the sampling interval, which in turn is equal to $\lambda_\mathrm{min}=2\Delta_x$, where $\Delta_x$ is the distance between two sampled points. If we now plug these values, related to different discretization levels, into Eq.~\eqref{eq:opt_t1} for a specific sliding velocity, the correspondent maximum durations of the normal loading phase to be taken into account, that minimizes the transient during the sliding process can be obtained. For the chosen discretization level of $m_x = 256$, a correspondent value of approximately $T_1 = 2.0\times10^{-6}\,\si{\second}$ is obtained and employed in the simulations that follow. It has to be remarked that different sliding velocities lead to different $T_1$. For the sake of simplicity the more demanding values, obtained for the highest velocity employed will be considered in all the simulations. Higher values of $T_1$, related to different possible discretization levels and sliding velocities, will not be taken into account.

\subsection{Additional model parameters}

A range of different sliding velocities is applied during Phase II to all the selected profiles. For each of them, twenty different values have been chosen, logarithmically centered around $v_0 = 10^ 4\,\si{\milli\meter/\second}$, and bounded, by $\vmin = 10^3\,\si{\milli\meter/\second}$ and $\vmax = 10^5\,\si{\milli\meter/\second}$. The intensity of $v_0$ has been chosen to approximately resemble the standard sliding velocity usually employed in the experimental British Pendulum friction tests (BPT)~\cite{liu:2003}, while the bounding values of the range have been designed to be far enough from the critical speed in order to clearly assess the maximum value of $\omu$ and its sensitivity to the changes in skid's velocity. A sliding length coincident with the complete development of the available profiles has been used in every case analyzed, leading to an overall value sliding length of $L \approx 90\,\si{\milli\meter}$, resulting, in turn, in a ratio with the skid length of $L/b \approx 9$. 
It has to be remarked that if a non-periodic profile is involved, the sliding length plays an important role in the determination of the average coefficient of friction since it has to be long enough for all the relevant roughness features to be taken into consideration. Finally, a value $p_0 = 2.0\,\si{\mega\pascal}$ for the vertical applied pressure has been employed, equal for all the three different profiles analyzed. Even though theoretical and experimental evidences~\cite{persson:1997,fortunato:2015} for its concurrent influence on the resulting viscoelastic coefficient of friction exist, its analysis goes beyond this preliminary study and it is therefore left for further investigations.

\subsection{Results and discussion}

\begin{figure}[b]
\centering
    \subfloat[][Instant value of the viscoelastic coefficient of friction for three different profiles in correspondence of the critical value of the sliding velocity. 
    \label{fig:mut}]
    {\includegraphics[width=0.45\textwidth]{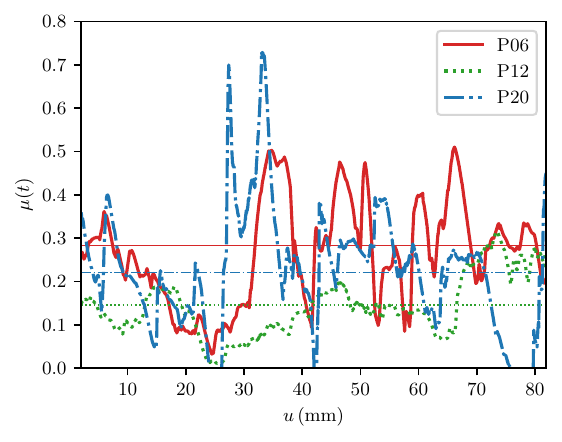}} \hspace{5mm}
    \subfloat[][Average value of the viscoelastic coefficient of friction, $\omu$ for every sliding velocity analyzed. For each profile a critical value close to the sliding velocity employed in the \emph{BTP} test can be observed.\label{fig:muav}]
    {\includegraphics[width=0.465\textwidth]{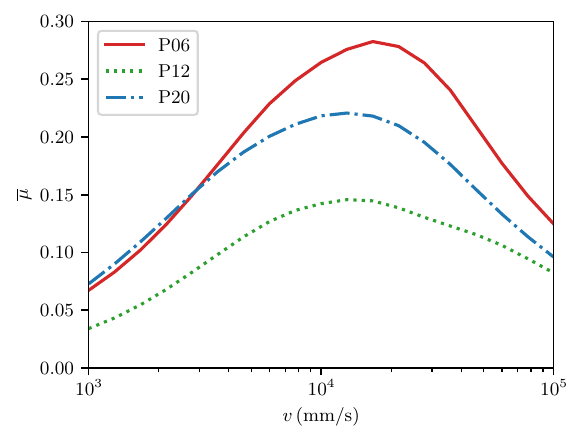}}
\caption{Simulation results in terms of viscoelastic coefficient of friction.}
\label{fig:final_results}
\end{figure}

Simulation results are exposed in Fig.~\ref{fig:final_results}. Figure~\ref{fig:mut} shows the instant values of viscoelastic coefficient of friction for the three different profiles in correspondence of the velocity value which determines the highest $\omu$. It can be noticed how the average values are in line with mean values obtained in literature, cfr., e.g.,~\cite{peng:2019}. Figure~\ref{fig:muav} shows the average values of the viscoelastic coefficient of friction $\omu$ for every sliding velocity analyzed and for each of the three profiles. In every case, a critical value for the velocity range can be clearly identified, which starts around $10\,\si{\meter/\second}$ for P12 and then slightly increases for P20 and P06.

Profile roughness parameters analysis exposed in Sec.~\ref{sec:roughness} showed that the profile labeled as P20 has the highest values for all the texture parameters considered. The surface from which this very same profile was extracted also shows the highest frictional performance measured by the SRT device. Even though this feature is not reproduced in the numerical simulation in terms of $\omu$ for most of the velocities under consideration, cfr. Fig.~\ref{fig:muav}, the profile is remarkably characterized by the highest instantaneous values of $\mu(t)$, as clearly visible from Fig.~\ref{fig:mut}, a response localized around $u \approx 30\,\si{\milli\meter}$.

Profile P12 has the lowest roughness parameters values, however the surface from which the profile was sectioned did not share this characteristic, i.e., was not characterized by the lowest friction performance determined by SRT measurements. On the other hand, for what concerns profile's parameters, the trend is well captured by the numerical framework, which delivers the lowest values of $\omu$ for all the velocities under consideration, and also the lowest values of maximum $\mu(t)$ for the sliding velocities considered.

For P06, good accordance is found between the SRT values related to the surface and the profile roughness parameters, since they are both characterized by the lowest indicators. 
Unfortunately, this trend is not captured at the simulation stage, which delivers unexpectedly high $\omu$ results. In this case the simulation was not able to capture the expected low average values for the hysteretic coefficient of friction. A possible explanation is related to missing features of the numerical model employed which is characterized by small displacements, linear material assumptions, an interface model which is not comprehensive of adhesion and Coulomb friction and a 2D setting.


Furthermore, a comparison between profile roughness parameters and measured SRT surface values is not straightforward, as the SRT values are related to the measured surface area and not to a single profile. As the choice of the profiles extracted from the digital surface models was arbitrary, it was not guaranteed that the measured frictional performance would follow the same trend as the roughness parameters values for the single profile. Pavement surfaces were evaluated for the friction performance to gain a general representation of surface roughness characteristics and to distinguish between the surfaces having significantly different friction performance. This was the key for the selection of surfaces for rough profiles extraction, which were later used in numerical simulations. Concerning this discrepancy, an explanation can be provided invoking missing additional mechanical effects related to the viscoelasticity of the bulk that, in conjunction with surface roughness, gives rise to a trend that cannot be explained according to the geometrical features of profile roughness alone.

\section{Conclusion}\label{sec:conc}

This research has extended the MPJR approach to address the complexities of tire-pavement interaction, by simulating the hysteretic frictional response of real pavement texture roughness features. By embedding detailed pavement profile models into the finite element method, the study avoids the need of geometrical approximations, maintaining high fidelity in representing profile interactions. The MPJR framework, previously validated for various contact problems, proves to be robust and reliable in capturing the characteristics of finite sliding scenarios, as highlighted by the employment of both simple sinusoidal and complex real rough profiles.

The findings of the study suggest that the MPJR approach is accurate and practical for evaluating the hysteretic coefficient of friction, delivering values which are in line with the literature in terms of magnitude. In spite a perfect correlation with specific roughness parameters derived from the analysis of the profiles could not be identified, this capability opens new possibilities for using the MPJR method as a consistent investigative tool in the field of tribology for applications requiring adequate modeling of rough contact interactions. The additional features provided by the current model are a necessary building block required to derive a comprehensive framework capable of delivering a better correlation with surface roughness parameters.

Profile-related roughness parameters calculated from the extracted pavement digital surface models showed to have a certain discrepancy with the experimentally measured friction performance, determined by the SRT device. 
The surface roughness evaluated by SRT device was approximated with a single 2D profile and its roughness features, generalizing the overall roughness of inspected surface. A better coincidence between calculated roughness parameters and experimentally validated skid resistance might be possible in case of 3D surface roughness parameters calculation from pavement digital surface model, which is planned in further research of pavement friction investigations.

After this preliminary study, future work will build on these results by incorporating large-scale 3D analyses, intuitively based on surface interpolation; more complex deformation scenarios and interface friction laws; and thermal effects. Moreover, the setting is prone to be embedded in extended analysis tools like time-series synthetic generation to artificially recreate wider surfaces for considering wider sliding ranges, thus leveraging on a surrogate modeling approach that would relieve expensive experimental data acquisition campaigns. Furthermore, model order reduction (MOR) techniques could also be put in place to reduce the high computational costs required by 3D simulations, also leveraging on HPC resources. All these new features will further consolidate the MPJR approach use for advanced frictional performance studies.

\appendix


\bibliographystyle{elsarticle-num} 
\bibliography{cas-refs}

\begin{thebibliography}{10}
\expandafter\ifx\csname url\endcsname\relax
  \def\url#1{\texttt{#1}}\fi
\expandafter\ifx\csname urlprefix\endcsname\relax\def\urlprefix{URL }\fi
\expandafter\ifx\csname href\endcsname\relax
  \def\href#1#2{#2} \def\path#1{#1}\fi

\bibitem{hall:2009}
N.~A. of~Sciences~Engineering, Medicine,
  \href{https://doi.org/10.17226/23038}{Guide for Pavement Friction} (2009).
\newline\urlprefix\url{https://doi.org/10.17226/23038}

\bibitem{bowden:1950}
F.~P. Bowden, D.~Tabor, \href{https://doi.org/10.1098/rspa.1939.0005}{The area
  of contact between stationary and moving surfaces}, Proceedings of the Royal
  Society of London. Series A. Mathematical and Physical Sciences 169~(938)
  (1939) 391--413.
\newblock \href {https://doi.org/10.1098/rspa.1939.0005}
  {\path{doi:10.1098/rspa.1939.0005}}.
\newline\urlprefix\url{https://doi.org/10.1098/rspa.1939.0005}

\bibitem{Popov:2010}
L.~V. Popov,
  \href{https://link.springer.com/book/10.1007/978-3-662-53081-8}{Contact
  Mechanics and Friction: Physical Principles and Applications}, Springer
  Heidelberg, 2010.
\newline\urlprefix\url{https://link.springer.com/book/10.1007/978-3-662-53081-8}

\bibitem{yu:2004}
N.~Yu, A.~A. Polycarpou,
  \href{https://linkinghub.elsevier.com/retrieve/pii/S0021979704005454}{Adhesive
  contact based on the {Lennard}–{Jones} potential: a correction to the value
  of the equilibrium distance as used in the potential}, Journal of Colloid and
  Interface Science 278~(2) (2004) 428--435.
\newblock \href {https://doi.org/10.1016/j.jcis.2004.06.029}
  {\path{doi:10.1016/j.jcis.2004.06.029}}.
\newline\urlprefix\url{https://linkinghub.elsevier.com/retrieve/pii/S0021979704005454}

\bibitem{sauer:2009}
R.~A. Sauer, P.~Wriggers,
  \href{https://linkinghub.elsevier.com/retrieve/pii/S0045782509002631}{Formulation
  and analysis of a three-dimensional finite element implementation for
  adhesive contact at the nanoscale}, Computer Methods in Applied Mechanics and
  Engineering 198~(49-52) (2009) 3871--3883.
\newblock \href {https://doi.org/10.1016/j.cma.2009.08.019}
  {\path{doi:10.1016/j.cma.2009.08.019}}.
\newline\urlprefix\url{https://linkinghub.elsevier.com/retrieve/pii/S0045782509002631}

\bibitem{wriggers:2009}
P.~Wriggers, J.~Reinelt,
  \href{https://linkinghub.elsevier.com/retrieve/pii/S0045782509000267}{Multi-scale
  approach for frictional contact of elastomers on rough rigid surfaces},
  Computer Methods in Applied Mechanics and Engineering 198~(21-26) (2009)
  1996--2008.
\newblock \href {https://doi.org/10.1016/j.cma.2008.12.021}
  {\path{doi:10.1016/j.cma.2008.12.021}}.
\newline\urlprefix\url{https://linkinghub.elsevier.com/retrieve/pii/S0045782509000267}

\bibitem{paggi&hills:2020}
M.~Paggi, D.~Hills,
  \href{https://link.springer.com/book/10.1007/978-3-030-20377-1}{Modeling and
  Simulation of Tribological Problems in Technology}, Springer, 2020.
\newline\urlprefix\url{https://link.springer.com/book/10.1007/978-3-030-20377-1}

\bibitem{zienkiewicz:2002}
O.~C. Zienkiewicz, R.~L. Taylor, The {Finite} {Element} {Method} - {Volume} 1:
  {The} basis, 5th Edition, Butterworth-Heinemann, Oxford, 2002.

\bibitem{brebbia:2017}
C.~A. Brebbia,
  \href{https://linkinghub.elsevier.com/retrieve/pii/S0955799716304453}{The
  birth of the boundary element method from conception to application},
  Engineering Analysis with Boundary Elements 77 (2017) iii--x.
\newblock \href {https://doi.org/10.1016/j.enganabound.2016.12.001}
  {\path{doi:10.1016/j.enganabound.2016.12.001}}.
\newline\urlprefix\url{https://linkinghub.elsevier.com/retrieve/pii/S0955799716304453}

\bibitem{peng:2019}
P.~Yi, J.~Q. Li, Y.~Zhan, K.~C.~P. Wang, G.~Yang,
  \href{https://doi.org/10.3390/ma12233821}{Finite element method-based skid
  resistance simulation using in-situ 3d pavement surface texture and friction
  data}, Materials 12 (23) (2019) 3821.
\newline\urlprefix\url{https://doi.org/10.3390/ma12233821}

\bibitem{yu:2020}
M.~Yu, Z.~You, G.~Wu, L.~Kong, C.~Liu, J.~Gao,
  \href{https://doi.org/10.1016/j.conbuildmat.2020.119878}{Measurement and
  modeling of skid resistance of asphalt pavement: A review}, Construction and
  Building Materials 260 (2020) 119878.
\newline\urlprefix\url{https://doi.org/10.1016/j.conbuildmat.2020.119878}

\bibitem{yu:2017}
M.~Yu, G.~Wu, L.~Kong, Y.~Tang,
  \href{https://doi.org/10.3390/app7111123}{Tire-pavement friction
  characteristics with elastic properties of asphalt pavements}, Applied
  Sciences 7 (11) (2017) 1123.
\newline\urlprefix\url{https://doi.org/10.3390/app7111123}

\bibitem{fwa:2017}
T.~Fwa, \href{https://doi.org/10.1016/j.ijtst.2017.08.001.}{Skid resistance
  determination for pavement management and wet-weather road safety},
  International Journal of Transportation Science and Technology 6, 3 (2017)
  217--227.
\newline\urlprefix\url{https://doi.org/10.1016/j.ijtst.2017.08.001.}

\bibitem{srirangam:2017}
S.~K. Srirangam, K.~Anupam, C.~Kasbergen, A.~T. Scarpas,
  \href{https://doi.org/10.1016/j.jtte.2017.07.004}{Analysis of asphalt mix
  surface-tread rubber interaction by using finite element method}, Journal of
  Traffic and Transportation Engineering (English Edition) Volume 4, Issue 4
  (2017) 395--402.
\newline\urlprefix\url{https://doi.org/10.1016/j.jtte.2017.07.004}

\bibitem{wagner:2017}
P.~Wagner, P.~Wriggers, L.~Veltmaat, H.~Clasen, C.~Prange, B.~Wies,
  \href{https://doi.org/10.1016/j.triboint.2017.03.015}{Numerical multiscale
  modelling and experimental validation of low speed rubber friction on rough
  road surfaces including hysteretic and adhesive effects}, Tribology
  International 111 (2017) 243--253.
\newline\urlprefix\url{https://doi.org/10.1016/j.triboint.2017.03.015}

\bibitem{tang:2018}
T.~Tang, K.~Anupam, C.~Kasbergen, R.~Kogbara, A.~S.~E. Masad,
  \href{https://doi.org/10.1177/0361198118796728}{Finite element studies of
  skid resistance under hot weather condition}, Transportation Research Record
  2672(40) (2018) 382--394.
\newline\urlprefix\url{https://doi.org/10.1177/0361198118796728}

\bibitem{oden:1985}
J.~Oden, J.~Martins, \href{https://doi.org/10.1016/0045-7825(85)90009-X}{Models
  and computational methods for dynamic friction phenomena}, Computer Methods
  in Applied Mechanics and Engineering 52 (1-3) (1985) 527--634.
\newline\urlprefix\url{https://doi.org/10.1016/0045-7825(85)90009-X}

\bibitem{liu:2019}
X.~Liu, Q.~Cao, H.~Wang, J.~Chen, X.~Huang,
  \href{https://doi.org/10.1177/0361198119832886}{Valuation of vehicle braking
  performance on wet pavement surface using an integrated tire-vehicle modeling
  approach}, Transportation Research Record 2673(3) (2019) 295--307.
\newline\urlprefix\url{https://doi.org/10.1177/0361198119832886}

\bibitem{persson:1997}
B.~N.~J. Persson,
  \href{https://meridian.allenpress.com/rct/article/70/1/1/92419/Hysteresis-Friction-of-Sliding-Rubbers-on-Rough?utm_source=chatgpt.com}{Hysteresis
  friction of sliding rubbers on rough and fractal surfaces}, Rubber Chemistry
  and Technology 70~(1) (1997) 1--55.
\newline\urlprefix\url{https://meridian.allenpress.com/rct/article/70/1/1/92419/Hysteresis-Friction-of-Sliding-Rubbers-on-Rough?utm_source=chatgpt.com}

\bibitem{paggi:2018}
M.~Paggi, J.~Reinoso, A variational approach with embedded roughness for
  adhesive contact problems, Mechanics of Advanced Materials and Structures
  27~(20) (2020) 1731--1747.
\newblock \href {https://doi.org/10.1080/15376494.2018.1525454}
  {\path{doi:10.1080/15376494.2018.1525454}}.

\bibitem{bonari:2021}
J.~Bonari, M.~Paggi, J.~Reinoso,
  \href{https://linkinghub.elsevier.com/retrieve/pii/S0168874X21000895}{A
  framework for the analysis of fully coupled normal and tangential contact
  problems with complex interfaces}, Finite Elements in Analysis and Design 196
  (2021) 103605.
\newblock \href {https://doi.org/10.1016/j.finel.2021.103605}
  {\path{doi:10.1016/j.finel.2021.103605}}.
\newline\urlprefix\url{https://linkinghub.elsevier.com/retrieve/pii/S0168874X21000895}

\bibitem{bonari:2020}
J.~Bonari, M.~Paggi,
  \href{https://www.mdpi.com/2075-4442/8/12/107}{Viscoelastic {Effects} during
  {Tangential} {Contact} {Analyzed} by a {Novel} {Finite} {Element} {Approach}
  with {Embedded} {Interface} {Profiles}}, Lubricants 8~(12) (2020) 107.
\newblock \href {https://doi.org/10.3390/lubricants8120107}
  {\path{doi:10.3390/lubricants8120107}}.
\newline\urlprefix\url{https://www.mdpi.com/2075-4442/8/12/107}

\bibitem{bonari:2022}
J.~Bonari, M.~Paggi, D.~Dini,
  \href{https://linkinghub.elsevier.com/retrieve/pii/S0020768322001640}{A new
  finite element paradigm to solve contact problems with roughness},
  International Journal of Solids and Structures 253 (2022) 111643.
\newblock \href {https://doi.org/10.1016/j.ijsolstr.2022.111643}
  {\path{doi:10.1016/j.ijsolstr.2022.111643}}.
\newline\urlprefix\url{https://linkinghub.elsevier.com/retrieve/pii/S0020768322001640}

\bibitem{mueser:2017}
M.~H. Müser, W.~B. Dapp, R.~Bugnicourt, P.~Sainsot, N.~Lesaffre, T.~A.
  Lubrecht, B.~N.~J. Persson, K.~Harris, A.~Bennett, K.~Schulze, S.~Rohde,
  P.~Ifju, W.~G. Sawyer, T.~Angelini, H.~Ashtari~Esfahani, M.~Kadkhodaei,
  S.~Akbarzadeh, J.-J. Wu, G.~Vorlaufer, A.~Vernes, S.~Solhjoo, A.~I. Vakis,
  R.~L. Jackson, Y.~Xu, J.~Streator, A.~Rostami, D.~Dini, S.~Medina,
  G.~Carbone, F.~Bottiglione, L.~Afferrante, J.~Monti, L.~Pastewka, M.~O.
  Robbins, J.~A. Greenwood,
  \href{http://link.springer.com/10.1007/s11249-017-0900-2}{Meeting the
  {Contact}-{Mechanics} {Challenge}}, Tribology Letters 65~(4) (2017) 118.
\newblock \href {https://doi.org/10.1007/s11249-017-0900-2}
  {\path{doi:10.1007/s11249-017-0900-2}}.
\newline\urlprefix\url{http://link.springer.com/10.1007/s11249-017-0900-2}

\bibitem{ban:2023}
I.~Ban, A model for skid resistance prediction based on non-standard pavement
  surface texture parameters, Ph.D. thesis, University of Rijeka Faculty of
  Civil Engineering (2023).

\bibitem{luhmann:2006}
T.Luhmann, S.Robson, S.Kyle, I.Harley, Close Range Photogrammetry:Principles,
  techniques and applications (2006).

\bibitem{bitelli:2012}
G.~Bitelli, A.~Simone, F.~Girardi, C.~Lantieri,
  \href{https://doi.org/10.3390/s120709110}{Laser scanning on road pavements: A
  new approach for characterizing surface texture}, Sensors 12 (2012)
  9110--9128.
\newline\urlprefix\url{https://doi.org/10.3390/s120709110}

\bibitem{callai:2022}
S.~C. Callai, M.~D. Rose, P.~Tataranni, C.~Makoundou, C.~Sangiorgi, R.~Vaiana,
  \href{https://doi.org/10.3390/coatings12121905}{Microsurfacing pavement
  solutions with alternative aggregates and binders: A full surface texture
  characterization}, Coatings 12(12) (2022) 1905.
\newline\urlprefix\url{https://doi.org/10.3390/coatings12121905}

\bibitem{zuniga:2019}
N.~Zuniga-Garcia, J.~A. Prozzi,
  \href{https://doi.org/10.1177/0361198118821598}{High-definition field texture
  measurements for predicting pavement friction}, Transportation Research
  Record 2673(1) (2019) 246--260.
\newline\urlprefix\url{https://doi.org/10.1177/0361198118821598}

\bibitem{kogbara:2018}
R.~B. Kogbara, E.~A. Masad, D.~Woodward, P.~Millar,
  \href{https://doi.org/10.1016/j.conbuildmat.2018.01.102.}{Relating surface
  texture parameters from close range photogrammetry to grip-tester pavement
  friction measurements}, Construction and Building Materials 166 (2018)
  227--240.
\newline\urlprefix\url{https://doi.org/10.1016/j.conbuildmat.2018.01.102.}

\bibitem{delorenzis:2013}
L.~De~Lorenzis, P.~Wriggers,
  \href{https://linkinghub.elsevier.com/retrieve/pii/S0927025613002218}{Computational
  homogenization of rubber friction on rough rigid surfaces}, Computational
  Materials Science 77 (2013) 264--280.
\newblock \href {https://doi.org/10.1016/j.commatsci.2013.04.049}
  {\path{doi:10.1016/j.commatsci.2013.04.049}}.
\newline\urlprefix\url{https://linkinghub.elsevier.com/retrieve/pii/S0927025613002218}

\bibitem{paggi:2011a}
M.~Paggi, P.~Wriggers,
  \href{https://linkinghub.elsevier.com/retrieve/pii/S0927025610006956}{A
  nonlocal cohesive zone model for finite thickness interfaces – {Part} {I}:
  {Mathematical} formulation and validation with molecular dynamics},
  Computational Materials Science 50~(5) (2011) 1625--1633.
\newblock \href {https://doi.org/10.1016/j.commatsci.2010.12.024}
  {\path{doi:10.1016/j.commatsci.2010.12.024}}.
\newline\urlprefix\url{https://linkinghub.elsevier.com/retrieve/pii/S0927025610006956}

\bibitem{paggi:2011b}
M.~Paggi, P.~Wriggers,
  \href{https://linkinghub.elsevier.com/retrieve/pii/S0927025610006920}{A
  nonlocal cohesive zone model for finite thickness interfaces – {Part} {II}:
  {FE} implementation and application to polycrystalline materials},
  Computational Materials Science 50~(5) (2011) 1634--1643.
\newblock \href {https://doi.org/10.1016/j.commatsci.2010.12.021}
  {\path{doi:10.1016/j.commatsci.2010.12.021}}.
\newline\urlprefix\url{https://linkinghub.elsevier.com/retrieve/pii/S0927025610006920}

\bibitem{barber:2018}
J.~Barber, \href{http://link.springer.com/10.1007/978-3-319-70939-0}{Contact
  {Mechanics}}, Vol. 250 of Solid {Mechanics} and {Its} {Applications},
  Springer International Publishing, Cham, 2018.
\newblock \href {https://doi.org/10.1007/978-3-319-70939-0}
  {\path{doi:10.1007/978-3-319-70939-0}}.
\newline\urlprefix\url{http://link.springer.com/10.1007/978-3-319-70939-0}

\bibitem{liu:2003}
Y.~Liu, T.~F. Fwa, Y.~S. Choo,
  \href{https://api.semanticscholar.org/CorpusID:108936458}{Finite-element
  modeling of skid resistance test}, Journal of Transportation Engineering-asce
  129 (2003) 316--321.
\newline\urlprefix\url{https://api.semanticscholar.org/CorpusID:108936458}

\bibitem{fortunato:2015}
G.~Fortunato, R.~D. Lorenzetti, A.~Bellini, E.~Ciulli,
  \href{https://arxiv.org/abs/1512.01359?utm_source=chatgpt.com}{On the
  dependency of rubber friction on the normal force or load: Theory and
  experiment}, arXiv preprint (2015).
\newblock \href {http://arxiv.org/abs/1512.01359} {\path{arXiv:1512.01359}}.
\newline\urlprefix\url{https://arxiv.org/abs/1512.01359?utm_source=chatgpt.com}

\end{thebibliography}

\end{document}